\documentclass[epj]{svjour}
\usepackage{graphicx,color,amssymb}
\usepackage[numbers,sort&compress,merge]{natbib}

\begin{document}

\pagestyle{plain}
\title{Constraints on the MSSM from the Higgs Sector}
\subtitle{A pMSSM Study of Higgs Searches, $B^0_s \to \mu^+ \mu^-$ and Dark Matter Direct Detection}
\author{A. Arbey\inst{1}\inst{2}\inst{3}, 
        M. Battaglia\inst{2} \inst{4} \inst{5} \and 
        F. Mahmoudi\inst{2} \inst{6}
}                     
%
%

\institute{Universit\'e de Lyon, France; Universit\'e Lyon 1, CNRS/IN2P3, UMR5822 IPNL, 
F-69622~Villeurbanne Cedex, France \and
CERN, CH-1211 Geneva 23, Switzerland \and
Observatoire de Lyon, CNRS, UMR 5574; Ecole Normale Sup\'erieure de Lyon, 
F-69561 Saint-Genis Laval Cedex, France \and
Santa Cruz Institute of Particle Physics, University of California, Santa Cruz,
CA 95064, USA \and
Lawrence Berkeley National Laboratory, Berkeley, CA 94720, USA \and
Clermont Universit\'e, Universit\'e Blaise Pascal, CNRS/IN2P3, LPC, BP 10448, 63000 Clermont-Ferrand, France}
\date{}

\abstract{
We discuss the constraints on Supersymmetry in the Higgs sector arising from LHC searches, 
rare $B$ decays and dark matter direct detection experiments. We show that constraints 
derived on the mass of the lightest $h^0$ and the CP-odd $A^0$ bosons from these searches 
are covering a larger fraction of the SUSY parameter space compared to searches for strongly 
interacting supersymmetric particle partners. We discuss the implications of a mass determination 
for the lightest Higgs boson in the range 123 $< M_h <$ 127~GeV, inspired by the intriguing hints 
reported by the ATLAS and CMS collaborations, as well as those of a non-observation of the lightest 
Higgs boson for MSSM scenarios not excluded at the end of 2012 by LHC and direct dark matter searches 
and their implications on LHC SUSY searches.
} 
\maketitle

\section{Introduction}
\label{sec:1}

The search for Supersymmetry (SUSY) is the main focus of the studies of physics beyond the Standard Model (SM) at the LHC.
Having acquired the status of possibly the best motivated theory of new physics over the past decades, Supersymmetry in 
its minimal incarnation (MSSM) with R-parity conservation is widely regarded as the template for theories and models with 
a conserved quantum number, leading to a candidate for relic dark matter and distinctive experimental signatures with 
hadrons, leptons and missing transverse energy, $MET$.

In view of the negative results of the searches in channels with $MET$ conducted by the ATLAS and CMS experiments on more 
than 1~fb$^{-1}$ of statistics at 7~TeV~\cite{Chatrchyan:2011zy,Aad:2011ib,Aad:2011cw,11-010,11-011}, questions on the 
continuing viability of the MSSM have arisen. 
These have been driven in particular by the strong impact of the LHC exclusion bounds on highly constrained MSSM models, such 
as the CMSSM and mSUGRA with few free parameters. Studies considering more general 
models without implicit correlations between the masses of the supersymmetric particle partners, such as the 19-parameter 
phenomenological MSSM (pMSSM), have demonstrated that a wide phase space of solutions, compatible with flavour physics, 
low energy data and dark matter constraints and beyond the current sensitivity of the LHC experiments, 
exists~\cite{Conley:2011nn, Sekmen:2011cz, Arbey:2011un}. Even at the end of the current LHC run, with an anticipated integrated 
luminosity of order of 15~fb$^{-1}$ per experiment, many of these solutions will not be tested. These solutions are compatible 
with all present bounds and $\sim$30\% of them have values of the fine tuning parameter, according to the definition of 
Ref.~\cite{Perelstein:2007nx}, below 100. This will make not possible to falsify 
the MSSM as the model of new physics beyond the SM and the source of relic dark matter in the universe, if no missing $E_T$ signal 
will be observed at the LHC by the end of 2012.

However, there is an alternative path to tightly constrain and test the MSSM at the LHC, which involves the bounds on its Higgs 
sector. Searches for the Higgs boson are expected to either discover or exclude a SM-like neutral Higgs boson with mass in 
the range 114 $< M_H <$ 127~GeV, which represents the current combined LEP and LHC mass bounds~\cite{seminar,ATLAS-CONF-2011-163,11-032} 
and corresponds to the mass range expected for the MSSM light Higgs boson, $h^0$, as well as that indicated by electro-weak data 
for a SM-like boson, $H^0_{SM}$. Presently, both ATLAS and CMS have a sensitivity to a light neutral Higgs boson in this mass 
range comparable to production yields expected in the SM.
In the MSSM, the $h^0$ mass depends on radiative corrections generated by SUSY loops and its couplings can be shifted resulting 
in a suppression (or an enhancement) of the production cross sections and decay branching fractions for the channels most relevant 
for the results of the LHC Higgs searches, compared to the SM predictions. However, such a suppression implies phenomenology which 
can be studied or constrained in direct SUSY searches and in rare $B$ decays at the LHC as well as in direct dark matter detection 
experiments, by the end of 2012. 

In this paper we discuss these constraints and their implications on the MSSM from a study of the pMSSM with 19 free 
parameters~\cite{Djouadi:1998di},
where we assume that the lightest neutralino, $\chi^0_1$ is the lightest supersymmetric particle (LSP). We further assume that 
the neutralino LSP is the weakly interacting massive particle (WIMP) responsible for (at least part of) the dark matter in the 
universe and focus on two scenarios for the light $h^0$. First, we consider a light Higgs boson with 123 $< M_{h^{0}} <$ 
127~GeV, as possibly suggested by the intriguing hints in the preliminary results of the Higgs searches at the LHC with 
almost 5~fb$^{-1}$ of data~\cite{seminar,ATLAS-CONF-2011-163,11-032}. Then, we consider 
the exclusion of the Higgs boson to a rate three times below that predicted in the SM, by the end of 2012. This paper is organised 
as follows. In section~\ref{sec:2} we summarise the dependence of the Higgs mass and of the mechanisms of suppression of Higgs 
event yield at the LHC on the MSSM parameters, while section~\ref{sec:3} investigates the present and projected bounds from 
LHC and DM experiments. We present the scenarios compatible with these bounds in section~\ref{sec:4}, while 
section~\ref{sec:5} has the conclusions.

\section{SUSY Higgs bosons and pMSSM scans}
\label{sec:2}

The calculation of the lightest Higgs boson mass, $M_{h^0}$, in the MSSM is the most precise prediction for a 
particle in the SUSY theory. After accounting for radiative corrections, $M_{h^0} \lesssim$ 
135~GeV~\cite{Carena:1995wu,Heinemeyer:1998kz,Carena:2000dp}. 
It is tantalising that this bound falls in the mass range indicated by electro-weak fits and within the window left 
open by Higgs searches at LEP-2, Tevatron and the LHC. This makes its search the most readily available method to either 
confirm or falsify a firm SUSY prediction. While the discovery of a Higgs-like particle in this mass range would also 
be compatible with the SM only, without any SUSY extension, and its exclusion could be reconciled with the SM, the 
exclusion of the Higgs boson in this range would put an end to the MSSM as viable extension of the SM. At the same 
time the determination of the lightest Higgs mass would provide us with constraints on the MSSM parameters. In particular, 
a value of $M_h$ around 125~GeV either rules out, or severely constrains, several scenarios so far extensively used in 
the study of Supersymmetry, such as the so-called ``no-mixing'' and ``typical-mixing'' scenarios~\cite{cmssm-higgs}.

\begin{table}
\begin{center}
\begin{tabular}{|c|c|}
\hline
~~~~Parameter~~~~ & ~~~~~~~~Range~~~~~~~~ \\
\hline\hline
$\tan\beta$ & [1, 60]\\
\hline
$M_A$ & [50, 2000] \\
\hline
$M_1$ & [-2500, 2500] \\
\hline
$M_2$ & [-2500, 2500] \\
\hline
$M_3$ & [50, 2500] \\
\hline
$A_d=A_s=A_b$ & [-10000, 10000] \\
\hline
$A_u=A_c=A_t$ & [-10000, 10000] \\
\hline
$A_e=A_\mu=A_\tau$ & [-10000, 10000] \\
\hline
$\mu$ & [-3000, 3000] \\
\hline
$M_{\tilde{e}_L}=M_{\tilde{\mu}_L}$ & [50, 2500] \\
\hline
$M_{\tilde{e}_R}=M_{\tilde{\mu}_R}$ & [50, 2500] \\
\hline
$M_{\tilde{\tau}_L}$ & [50, 2500] \\
\hline
$M_{\tilde{\tau}_R}$ & [50, 2500] \\
\hline
$M_{\tilde{q}_{1L}}=M_{\tilde{q}_{2L}}$ & [50, 2500] \\
\hline
$M_{\tilde{q}_{3L}}$ & [50, 2500] \\
\hline
$M_{\tilde{u}_R}=M_{\tilde{c}_R}$ & [50, 2500] \\
\hline
$M_{\tilde{t}_R}$ & [50, 2500] \\
\hline
$M_{\tilde{d}_R}=M_{\tilde{s}_R}$ & [50, 2500] \\
\hline
$M_{\tilde{b}_R}$ & [50, 2500] \\
\hline
\end{tabular}
 \end{center}
\caption{pMSSM parameter ranges adopted in the scans (in GeV when applicable).\label{tab:paramSUSY}}
\end{table}
In order to study the properties of SUSY Higgs bosons, we perform a flat scans of the pMSSM, where we 
vary its 19 parameters in an uncorrelated way within the ranges given in Table~\ref{tab:paramSUSY}, and 
generate a total of 40M points. The scan range is explicitely chosen to include the so-called ``maximal mixing'' 
region~\cite{Carena:2002qg}, at $X_t \sim \sqrt{6}M_{\mathrm{SUSY}}$, where $X_t = A_t -\mu \cot \beta$ and $
M_{\mathrm{SUSY}} = \sqrt{m_{\tilde t_1} m_{\tilde t_2}}$, which corresponds to larger values of $M_{h^0}$.  
We select the set of points fulfilling constraints from flavour physics and lower energy 
searches at LEP-2 and the Tevatron, as discussed in~\cite{Arbey:2011un}. We find 65 $< M_{h^0} <$ 143~GeV 
for points compatible with those constraints . In addition, we perform dedicated scans for specific scenarios 
yielding large suppression of the Higgs production and decay channels explored at the LHC, where we restrict 
the ranges of a subset of the pMSSM parameters, as discussed in section~3.

The details of the pMSSM scans and the tools used for the computations of the spectra and relevant observables have been 
presented in details elsewhere~\cite{Arbey:2011un}. Here we mention only those most relevant to this study. 
SUSY mass spectra are generated with {\tt SOFTSUSY 3.2.3}~\cite{Allanach:2001kg}.  The decay branching fractions of Higgs 
bosons are obtained using {\tt HDECAY 4.40}~\cite{Djouadi:1997yw} including gaugino and sfermion loop corrections, and 
cross-checked with {\tt FeynHiggs 2.8.5}~\cite{Heinemeyer:1998yj,Heinemeyer:1998np}. The widths and decay branching 
fractions of the other SUSY particles are computed using {\tt SDECAY 1.3}~\cite{Muhlleitner:2003vg}. The dark matter relic 
density is calculated with {\tt SuperIso Relic v3.2}~\cite{Mahmoudi:2007vz,Mahmoudi:2008tp,Arbey:2009gu}, which provides us 
also with the flavour observables. The neutralino-nucleon scattering cross-sections are computed with 
{\tt micrOMEGAs 2.4} \cite{Belanger:2008sj}. The $gg$ and $bb$ Higgs production cross sections are computed using 
{\tt HIGLU 1.2}~\cite{Spira:1995rr,Spira:1996if} and {\tt FeynHiggs 2.8.5}, respectively. The Higgs production cross sections 
and the branching fractions for decays into $\gamma \gamma$ and $WW$, $ZZ$ from {\tt HIGLU} and {\tt HDECAY} are compared 
to those predicted by {\tt FeynHiggs}. In the SM both the $gg \to H^0_{SM}$ cross section and the branching fractions agree 
within  $\sim$3\%. Significant differences are observed in the SUSY case, with {\tt HDECAY} giving values of the branching 
fractions to $\gamma \gamma$ and $WW$, $ZZ$ which are on average 9\% lower and 19\% larger than those of {\tt FeynHiggs} and have 
an r.m.s.\ spread of the distribution of the relative difference between the two programs of  18\% and 24\%, respectively. 
In the following, we use {\tt HDECAY} throughout the analysis and we comment on the effects of using the results of 
{\tt FeynHiggs} instead at the end of section \ref{sec:4-1}.
Then, we test the sensitivity to the $b$-quark mass by varying the $b$ pole mass by $\pm$200~MeV. The corresponding relative 
change of the SUSY to SM ratio of the branching fractions is $\pm$0.1~\% for both $\gamma \gamma$ and $WW$/$ZZ$ decays. 
Finally, we test the decoupling limit by taking the ratio of the SUSY branching fractions to their SM counterparts for $M_A >$ 
700~GeV for {\tt HDECAY}. We find 0.994$\pm$0.005 and 0.954$\pm$0.005 for $\gamma \gamma$ and $WW$/$ZZ$. 
For computing the ratio of the $\gamma \gamma$ branching fraction in SUSY and the SM and test the decoupling limit, the 
electro-weak corrections to the SM $\gamma \gamma$ decay width~\cite{Passarino:2007fp}, which decrease the SM branching 
fraction by 2 to 3\% in the mass interval 90 $< M_H <$ 140~GeV and are not known for the MSSM, are removed.

\subsection{Coupling suppression through SUSY corrections}
\label{sec:2-1}

The ratios of the $h^0$ couplings to up-type quarks and gauge bosons to their SM values scale as 
$\cos \alpha / \sin \beta$ and $\sin(\beta - \alpha)$, respectively, where  $\alpha$ is the mixing 
angle in the CP-even Higgs sector and depends on $M_A$ for small to intermediate values of its mass.
This induces a decrease of the couplings to top and $W^{\pm}$, $Z^0$ which propagates to the 
$\sigma(g g \rightarrow h^0)$, $h^0 \rightarrow \gamma \gamma$, $W^+W^-$ and 
$h^0 \rightarrow \gamma \gamma$, $Z^0Z^0$ branching fractions.

At relatively high values of $\tan\beta$ and for some values of the soft SUSY parameters which enter in 
radiative corrections for the Higgs sector, the coupling of the $h^0$ boson to $b$ quarks and $\tau$ 
leptons becomes strongly suppressed. In the case of the $h^0b \bar b$ couplings, additional suppression 
could also occur as a result of vertex corrections involving gluino/sbottom loops. 

In the decoupling regime, scalar top and, to a lesser extent, bottom contributions may still suppress the 
gluonic width and thus the $g g \rightarrow h^0$ cross section as well as the $h^0 \rightarrow \gamma \gamma$ 
decay branching fraction. This is important for light squark masses and large values of the mixing parameter 
in the stop sector, $X_t$~\cite{Djouadi:1996pb, Djouadi:1998az, Djouadi:2005gj, Djouadi:2005zz}.

Finally, at high $\tan\beta$ and low $M_A$ values, there is the so-called
``intense coupling'' regime~\cite{Boos:2002ze,Boos:2002ze} in which the three
neutral Higgs bosons are very close in mass and have enhanced couplings to
bottom quarks and tau leptons and thus, reduced branching ratios to $W$ and $Z$ 
bosons and to photons. Some points of the parameter space in which this scenario 
occurs, in particular at $\tan \beta >10$,  should be removed by the 
$h^0, H^0, A^0 \to \tau \tau$ searches.

\subsection{Rate suppression through decays to SUSY particles}
\label{sec:2-2}

Decays into pairs of SUSY particle, with mass smaller than $M_{h^0}/2$, modify the $h^0$ branching 
fractions into SM particles. In practice there are two scenarios which are relevant to the reduction of 
the SM yields. If the LSP neutralino is light, the decay 
$h^0 \rightarrow \tilde\chi^0 \tilde\chi^0$ may induce a large $h^0$ invisible decay width and suppress 
its standard decays for some specific combinations of the $M_1$, $M_2$ and $\mu$ 
parameters~\cite{Djouadi:2001kba,Djouadi:2005gj,Djouadi:2005zz}. This scenario has received only some marginal 
attention~\cite{Cavalli:2002vs,Kinnunen:2005aq} so far{\footnote{It has now been reconsidered in details 
in~\cite{Vasquez:2011qc}.}}. It is now becoming an intriguing 
possibility in view of the results by the DAMA, CoGENT and CRESST experiments, all reporting excess of 
events compatible with the interaction of a light WIMP with large scattering cross section on nucleons.
We study this possibility with a dedicated scan where we restrict the parameter ranges to 
$-120 < M_1 < 120$ GeV and $-650 < M_2 < 650$ GeV. Detailed results on pMSSM scenarios compatible with 
the possible signal reported by the three experiments are given elsewhere~\cite{susyscan2}. Here, we 
do not consider the $\tilde \chi p$ cross section as a constraint and study instead the modification 
of the Higgs branching fractions induced by decays into $\tilde\chi^0_1$ pairs. 
\begin{figure}[bh!]
\begin{center}
\begin{tabular}{c}
\includegraphics[width=0.35\textwidth]{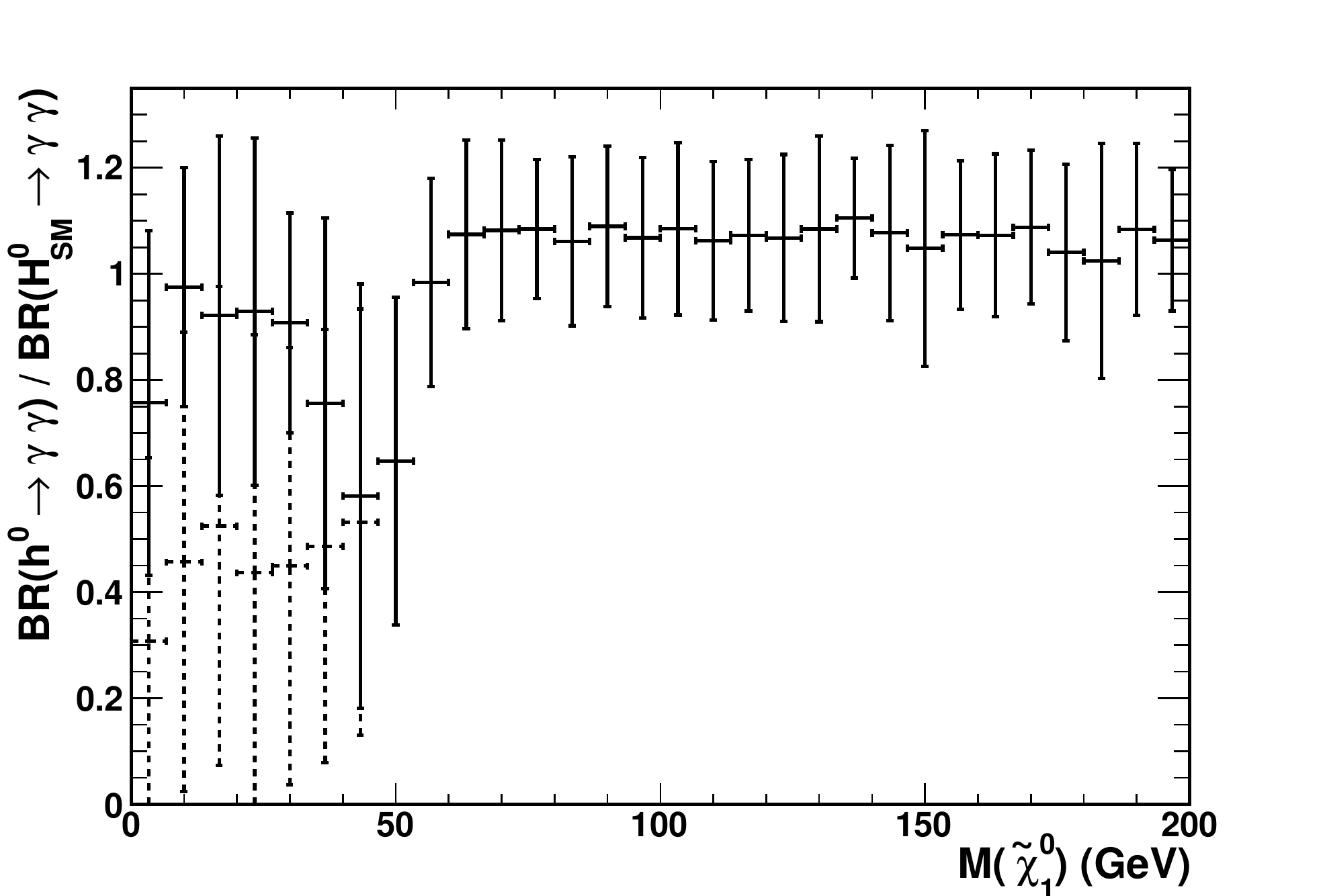} \\
\includegraphics[width=0.35\textwidth]{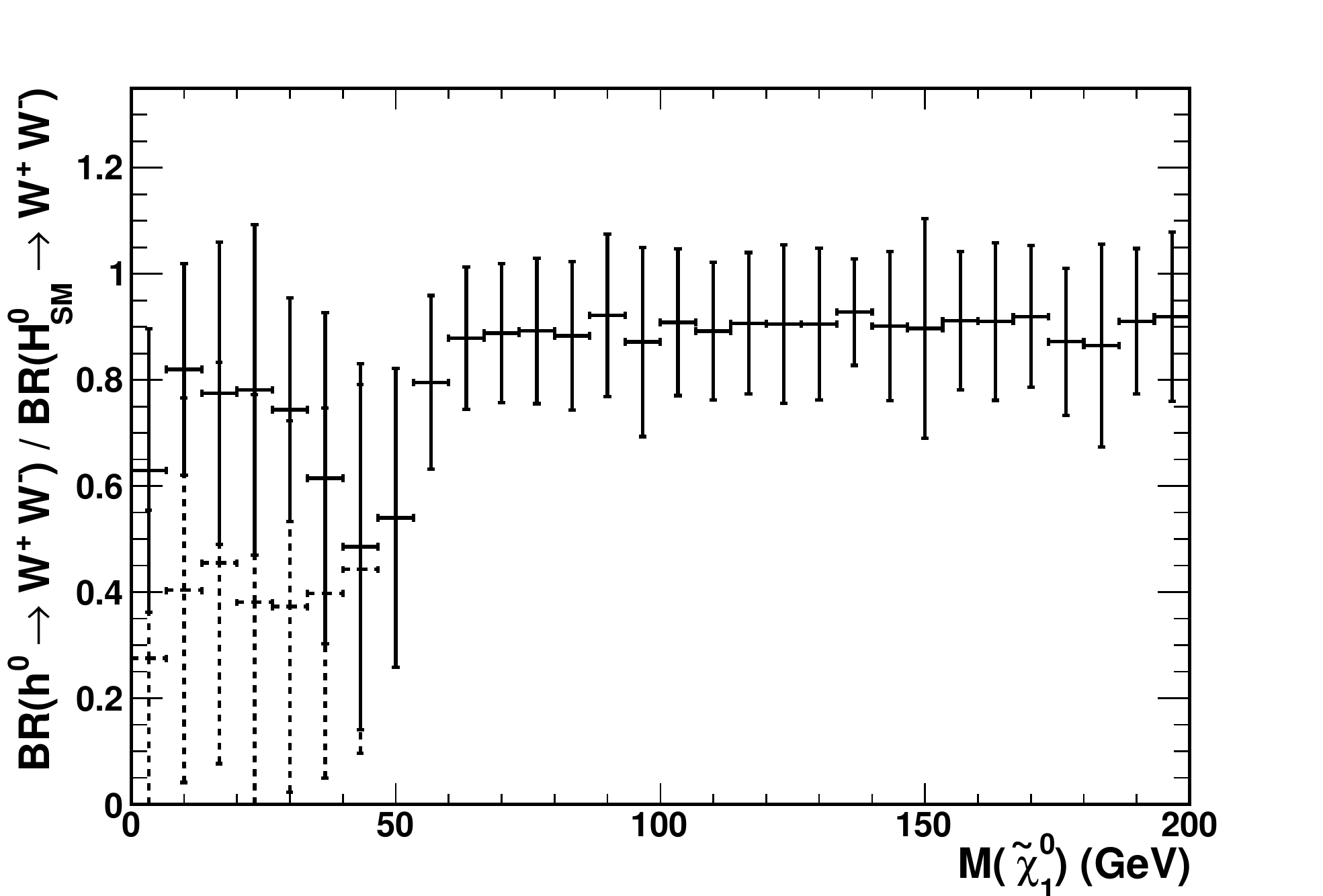}
\end{tabular}
\end{center}
\caption{The ratio of the branching fraction for $h^0 \rightarrow \gamma \gamma$ (upper panel) and 
$W^+ W^-$, $Z^0Z^0$ (lower panel) to the SM prediction, obtained with HDECAY, as a function of the lightest 
neutralino mass for pMSSM points with $A^0$ boson, $\tilde t_1$ and $\tilde b_1$ masses above 500~GeV. 
The dashed and full vertical bars give the full range of values for pMSSM points before and after applying 
the constraint on the $Z^0$ invisible width, 
respectively.}
\label{fig:hBRinv}
\end{figure}
The constraint from the $Z^0$ invisible decay width measured at LEP restricts the parameter space to 
points where the $\tilde\chi^0_1$ is bino-like, if its mass is below 45~GeV, and thus  to relatively large 
values of the higgsino mass parameter $|\mu|$. Since a large decay width into $\tilde \chi^0_1 \tilde \chi^0_1$ 
corresponds to small values of $|\mu|$, this remove a large part of the parameter space where the invisible 
Higgs decay width is large. Still, we observe an important suppression of decays into $\gamma \gamma$ and 
$W^+W^-$, $Z^0 Z^0$ for 45~GeV$< M_{\tilde\chi^0} < M_{h^0}/2$ and $| \mu | <$ 150, corresponding to a combination 
of parameters where the $\tilde\chi^0_1$ is a mixed higgsino-gaugino state. In this region, the suppression reaches 
values up to a factor of five, which may upset the sensitivity of the LHC light Higgs searches in the canonical 
channels (see Figure~\ref{fig:hBRinv}).
The second scenario has the lightest stau, $\tilde \tau^{\pm}_1$, below threshold for the 
$h^0 \rightarrow \tilde\tau^+_1 \tilde\tau^-_1$ decay. 
In this case the decay into stau pairs almost saturates the $h^0$ decay width and again suppresses 
the $\gamma \gamma$ and $WW$/$ZZ$ rates~\cite{Djouadi:1999xd}. Since the $M_{\tilde \tau_1} - M_{\tilde \chi^0_1}$ 
is small the $h^0$ decay consists of two very soft $\tau$ jets which would fail the trigger cuts at the LHC. 
However, this scenario is tightly constrained by the results of the LEP-2 searches.

\section{Current and projected bounds}
\label{sec:3}

We consider four sets of constraints on SUSY parameters. These are from direct searches i) for 
SUSY particles with $MET$ signatures and ii) for $A^0$ bosons in the channel 
$A^0 \rightarrow \tau^+ \tau^-$, iii) from the $B_s \rightarrow \mu^+ \mu^-$ rare decay and 
iv) from dark matter direct detection experiments. These constraints, originating from different 
sectors of the theory, are all sensitive to the SUSY parameters most relevant for setting the couplings 
and decay branching fractions of the light $h^0$ bosons. Their combination provides the boundary 
conditions for the parameter space where we test the possible suppression of the yields in the LHC 
Higgs searches.
We start from the situation outlined by the current data and project towards the status of these 
bounds at the time of the completion of the LHC run, for $\sqrt{s}$=7~TeV, at the end of 2012, assuming 
no signal is observed, except for the $B_s \rightarrow \mu^+ \mu^-$ decays, for which we assume a branching 
fraction equal to its SM expectation.

\subsection{Direct $\tilde g$ and $\tilde q$ searches at LHC}
\label{sec:3-1}

The direct searches for $MET$ signatures with jets and leptons probe $\tilde g$
and $\tilde q$ masses up to $\sim$500~GeV with 1~fb$^{-1}$ and $\sim$750~GeV with 
15~fb$^{-1}$ at 7~TeV~\cite{Conley:2011nn,Sekmen:2011cz,Arbey:2011un}. LHC operation 
at 8~TeV for an integrated luminosity of 15~fb$^{-1}$ will further push this sensitivity.
While these are significantly constraining its 
parameter space, they are hardly decisive in disproving the MSSM as a viable 
theory. In fact, the gluino and squark masses can be pushed beyond those
kinematically accessible in the current LHC run and still the MSSM would 
have all those features which have made SUSY so popular as SM extension, though 
at the cost of an increase of the fine tuning parameter.

\begin{figure}[h!]
\begin{center}
\begin{tabular}{c c}
\includegraphics[width=0.23\textwidth]{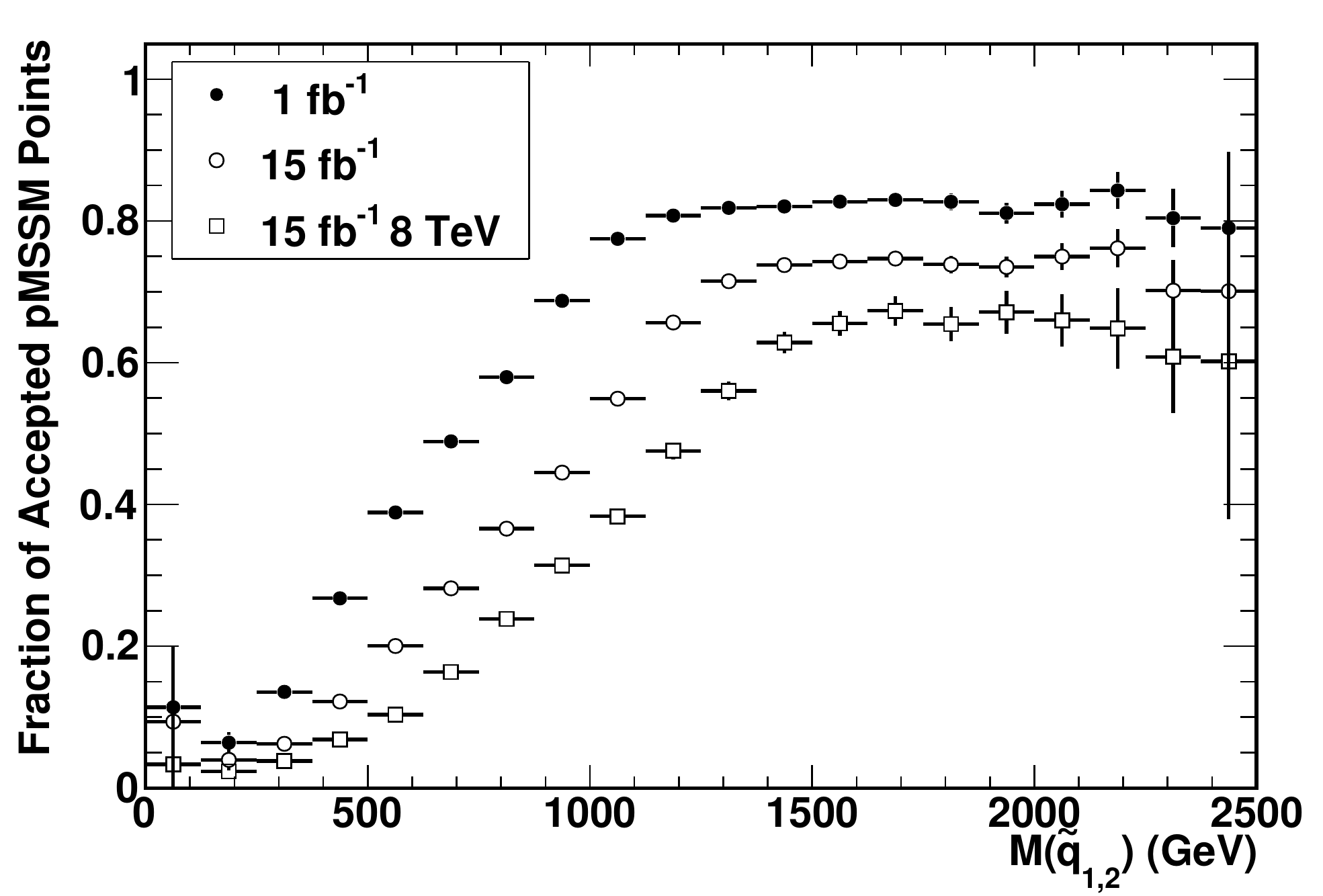} &
\includegraphics[width=0.23\textwidth]{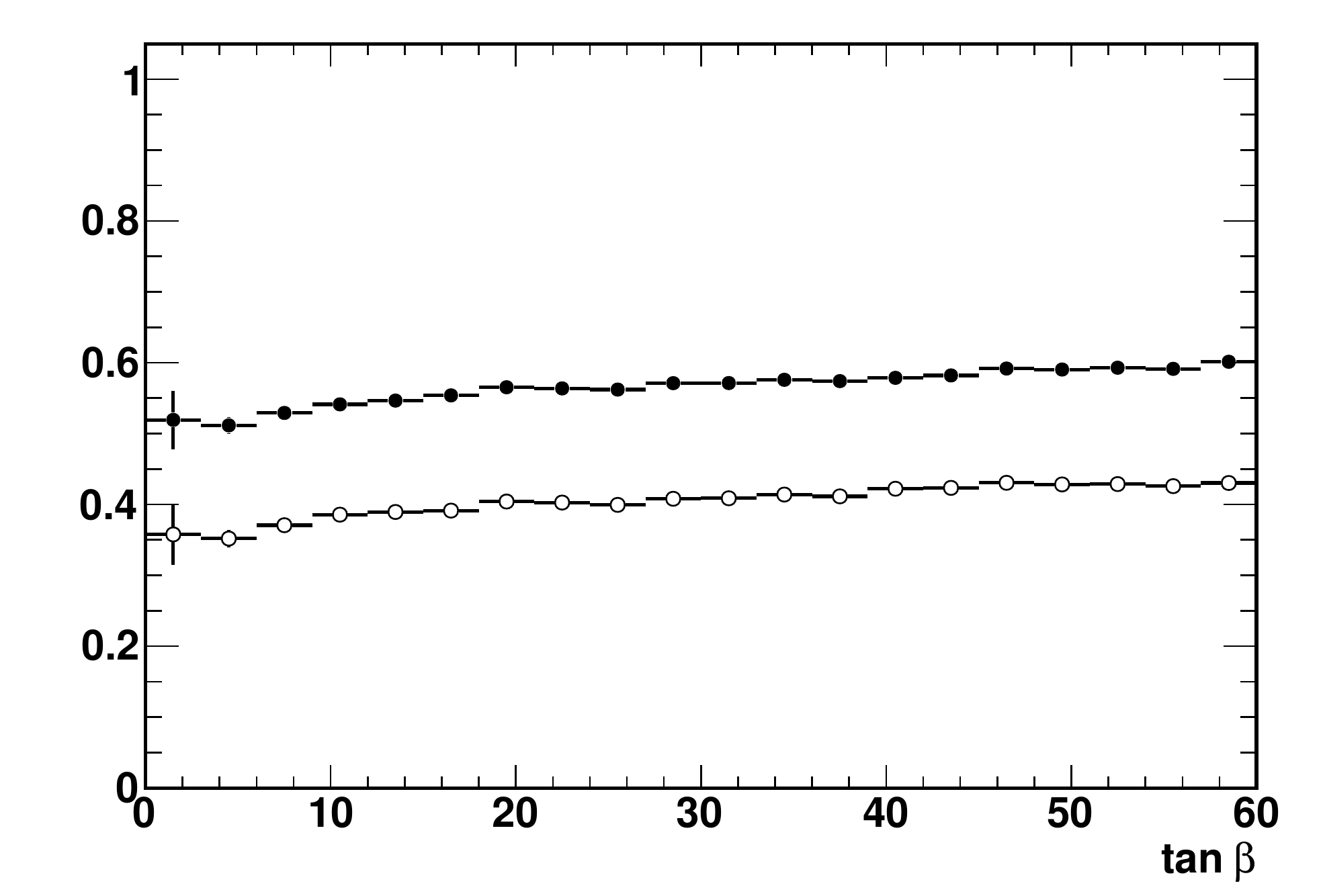} \\
\includegraphics[width=0.23\textwidth]{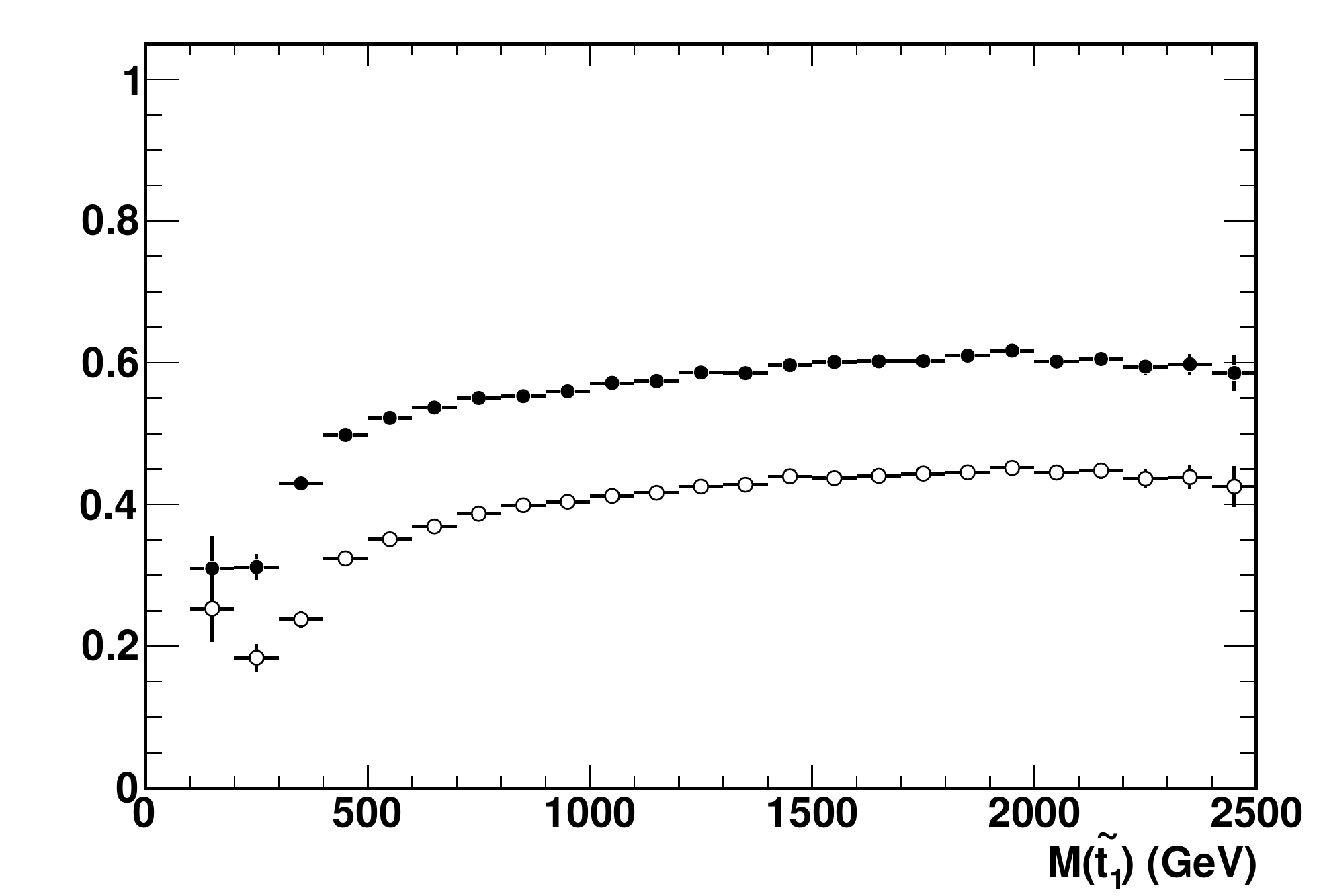} &
\includegraphics[width=0.23\textwidth]{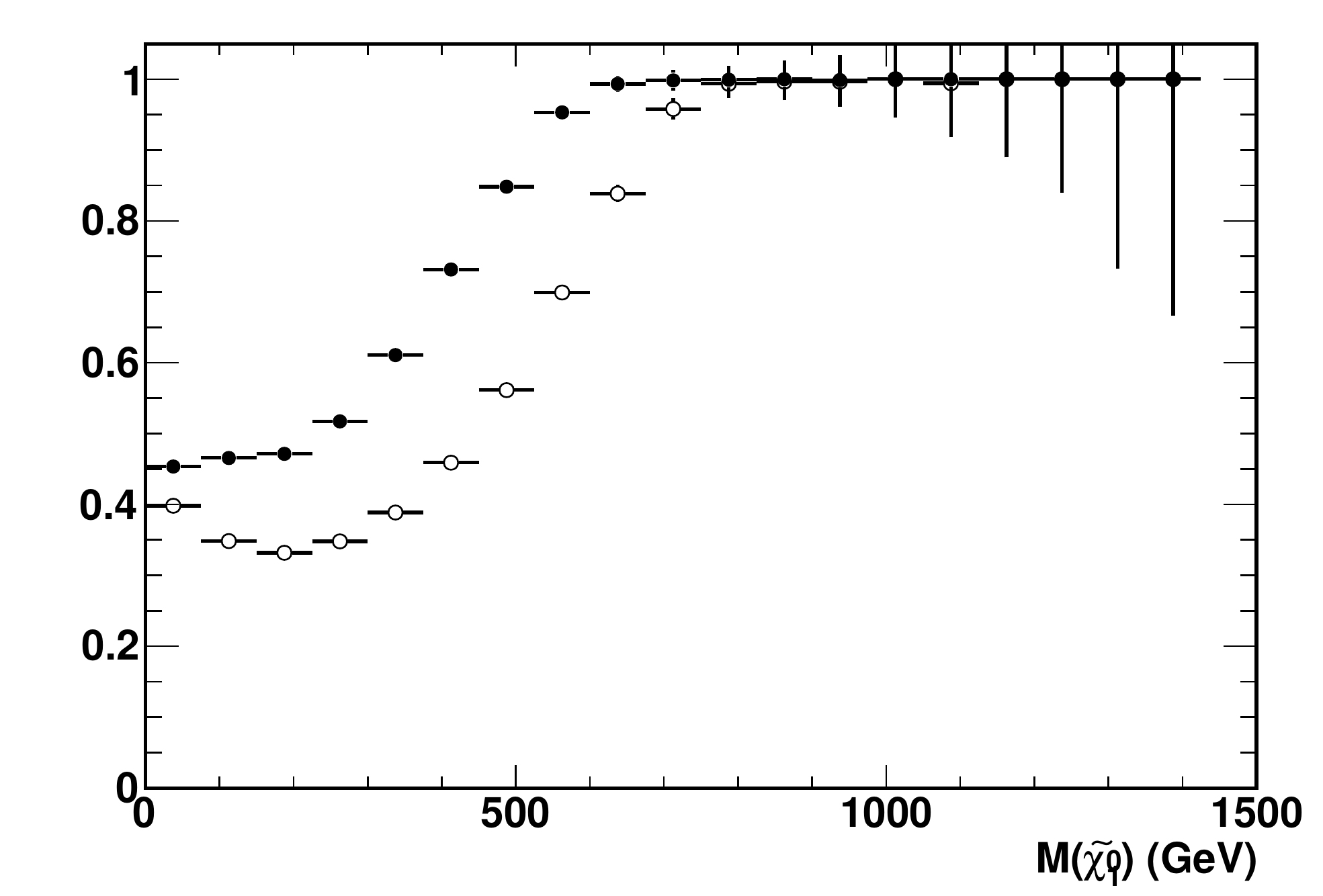} \\
\end{tabular}
\end{center}
\caption{Fraction of accepted pMSSM points not excluded by the SUSY searches on 1 (filled circles) and 15~fb$^{-1}$ 
of LHC data at 7~TeV (open circles) and at 8~TeV (open squares) as a function of the mass of the lightest squark of 
the first two generations (upper left panel), of the mass of the scalar top $\tilde t_1$ (lower left panel), of 
$\tan \beta$ (upper right panel) and of the mass of the lightest neutralino $\tilde \chi^0_1$ (lower right panel).}
\label{fig:st1b1}
\end{figure}

Of special interest, in relation to the Higgs sector, are the constraint derived 
on the mass of the lightest scalar top, $\tilde t_1$ quark. This squark is correlated 
with $M_h$ and it can play a role together with the $\tilde b_1$ in modifying the 
$h^0$ couplings. Figure~\ref{fig:st1b1} shows the fractions of accepted pMSSM points which 
are compatible with the results of the CMS analyses in the fully hadronic~\cite{Chatrchyan:2011zy} 
and in the leptonic channels~\cite{11-010,11-011} on 1~fb$^{-1}$ and its projection for 15~fb$^{-1}$ 
at 7~TeV, as a function of the masses of the lightest squark of the first two generations 
$\tilde q_{1,2}$ and of the $\tilde t_1$. These are obtained performing the same analysis as 
in~\cite{Arbey:2011un}. In the upper left panel of Figure~\ref{fig:st1b1} we also present a first 
estimate of the improvement of the sensitivity to scalar quarks of the first two generation for 
8~TeV LHC operation. 
From the results of generic scalar quark searches, which are not optimised for $\tilde t$, and dedicated $t \bar t$ + 
MET analyses, as that of ref.~\cite{Aad:2011wc}, 15~fb$^{-1}$ of  LHC data should be sensitive to MSSM solutions with 
light scalar quarks of the third generation with masses $\sim$300-400~GeV (see also~\cite{Papucci:2011wy}). Sensitivity 
beyond this 
mass limit is limited by the small production cross sections and the large backgrounds from top events. On the other hand, 
after removing pMSSM points excluded by the LHC searches, the acceptance w.r.t.\ other variables of interest here, such 
as $M_A$ and $\tan \beta$, is flat, indicating that the gluino and scalar quark searches do not influence the 
Higgs sector parameters.

\subsection{Direct $A^0 \rightarrow \tau^+ \tau^-$ searches at LHC}
\label{sec:3-2}

The result of the direct search for the $A^0$ boson at the LHC is the single most constraining 
piece of information on the ($M_A , \tan \beta$) plane. The CMS collaboration has presented the 
results of a search for neutral Higgs bosons decaying into $\tau$ pairs based on  1.1~fb$^{-1}$ 
of integrated luminosity~\cite{Chatrchyan:2011nx} and recently reported a preliminary update based 
on the analysis of 4.6~fb$^{-1}$~\cite{seminar}. 
The search not revealing any significant excess of events, limits on the product of production cross 
section and decay branching fraction as a function of the boson mass, corresponding to the 95\% C.L. 
expected bound are given in~\cite{11-009}.
\begin{figure}[t!]
\begin{center}
\begin{tabular}{c}
\includegraphics[width=0.35\textwidth]{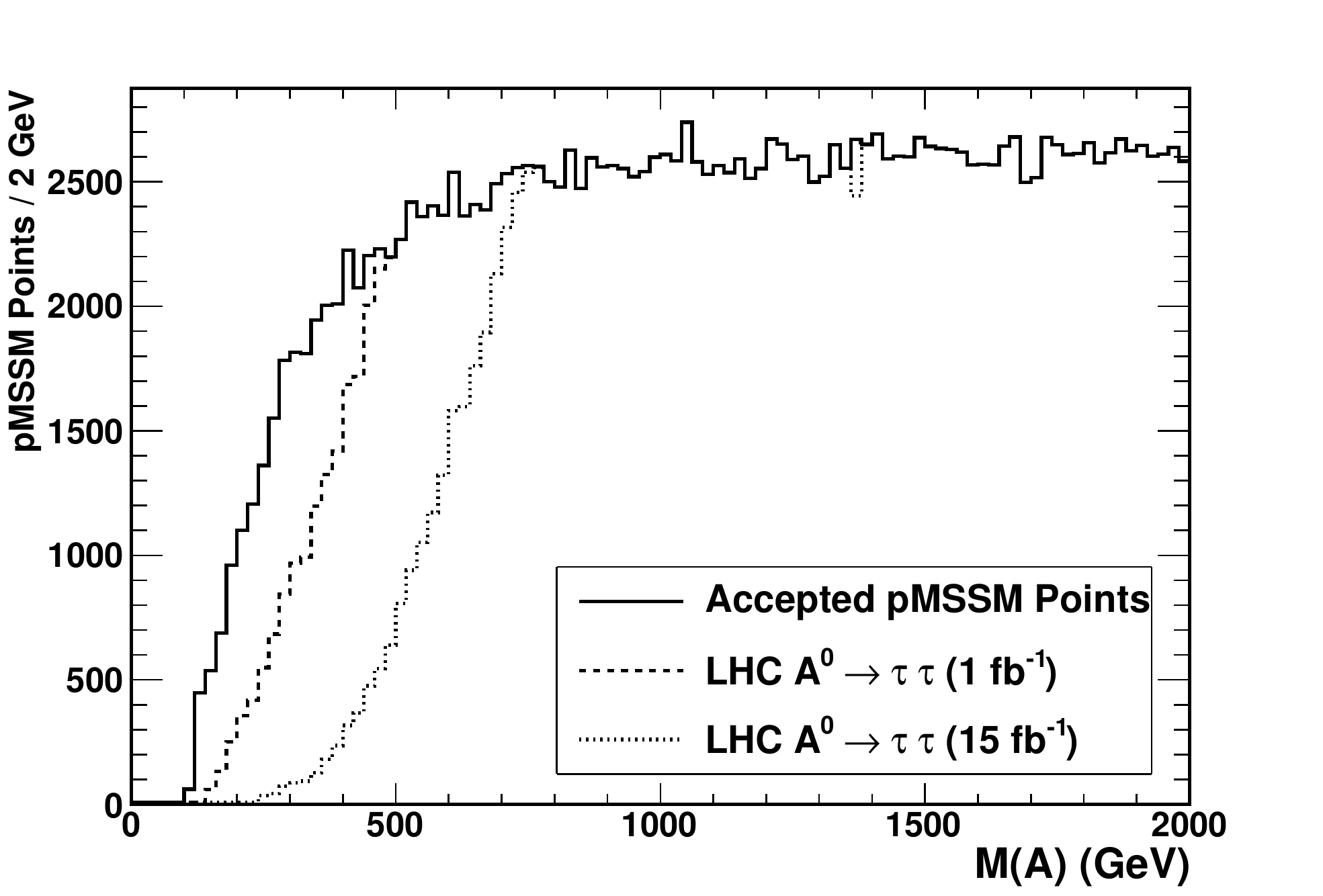} \\
\includegraphics[width=0.35\textwidth]{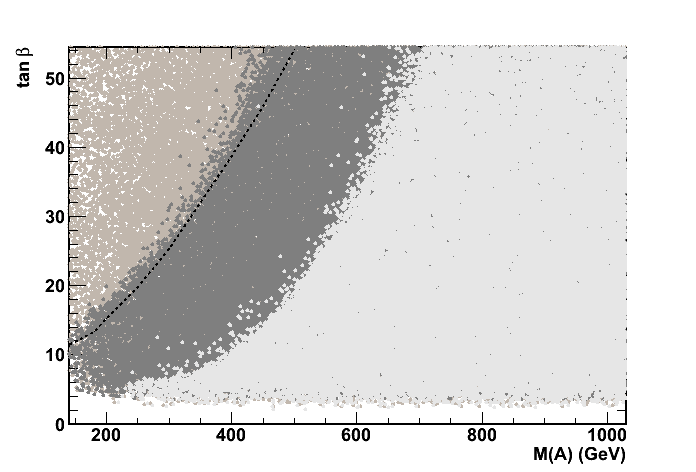} \\
\end{tabular}
\end{center}
\caption{Distribution of pMSSM points after the $A^0 \rightarrow \tau^+ \tau^-$ search projected on the 
$M_A$ (upper panel) and ($M_A , \tan \beta$) plane (lower panel) for all accepted pMSSM points (medium grey), 
points not excluded with 1~fb$^{-1}$ of data (dark grey) and the projection for the points not excluded 
with  15~fb$^{-1}$ of data (light grey). The dashed line on the ($M_A , \tan \beta$) plot indicates the 
95~\% C.L. limit derived by CMS in the $M_h$-max scenario with $M_{\mathrm{SUSY}}$ = 1~TeV for 1.1~fb$^{-1}$.}
\label{fig:atau}
\end{figure}
In order to map these bounds on the ($M_A , \tan \beta$) plane for the pMSSM and project them to 
15~fb$^{-1}$ of data, we compute the product of production cross section and decay branching fraction 
into $\tau$ pairs for the $A^0$ for each accepted pMSSM point. First we validate our procedure by taking 
the contour of the points having this product equal to that corresponding to the CMS expected limit. 
The right panel of Figure~\ref{fig:atau} shows this region for pMSSM points compared to the published CMS 
contour. These agree within 15\%. Then, we rescale the product to reproduce the projected limit for 
15~fb$^{-1}$ and remove the points which can be excluded if no signal is observed. Figure~\ref{fig:atau} 
shows the points surviving this selection in the $M_A$ and  ($M_A , \tan \beta$) parameter space. We note 
that the 2012 data should severely constrain the low $M_A$ scenario by removing all solutions with 
$M_A <$ 220~GeV and restricting the region with $M_A <$ 400~GeV to $\tan \beta$ values below 10. 
However, a tiny region with 220 $< M_A <$ 350~GeV survives for $\tan \beta \simeq$ 5. 

\subsection{$B^0_s \rightarrow \mu^+ \mu^-$ at LHC}
\label{sec:3-3}

The decay $B_s \to \mu^+ \mu^-$ is very sensitive to the presence of SUSY particles. 
At large $\tan\beta$, the SUSY contribution to this process is dominated by the exchange 
of neutral Higgs bosons, and very restrictive constraints can be obtained on the supersymmetric 
parameters~\cite{Akeroyd:2011kd}. Indeed, the couplings of the neutral Higgs bosons to $b$ quark 
and muons are proportional to $\tan\beta$, which can lead to enhancement of orders of magnitude 
compared to the SM value, which is helicity suppressed.

The $B_s \to \mu^+ \mu^-$ decay has been searched for at the Tevatron and the LHC. The CDF 
experiment has reported and excess of events corresponding to a branching fraction 
of ($1.8^{+1.1}_{-0.9}$)$\times$10$^{-8}$~\cite{PhysRevLett.107.191801}. The LHCb and CMS 
collaborations did not observe any significant excess and released a 95\% C.L.\ combined limit 
of $\mathrm{BR}(B_s\to\mu^+\mu^-) < 1.1 \times 10^{-8}$~\cite{CMS_plus_LHCb}, which is only 
$\sim$4 times above the SM predictions. In order to take into account the theoretical 
uncertainties, in our numerical analysis we adopt the limit 
$\mathrm{BR}(B_s\to\mu^+\mu^-) < 1.26 \times 10^{-8}$.

\begin{figure}[t!]
\begin{center}
\begin{tabular}{c}
\includegraphics[width=0.35\textwidth]{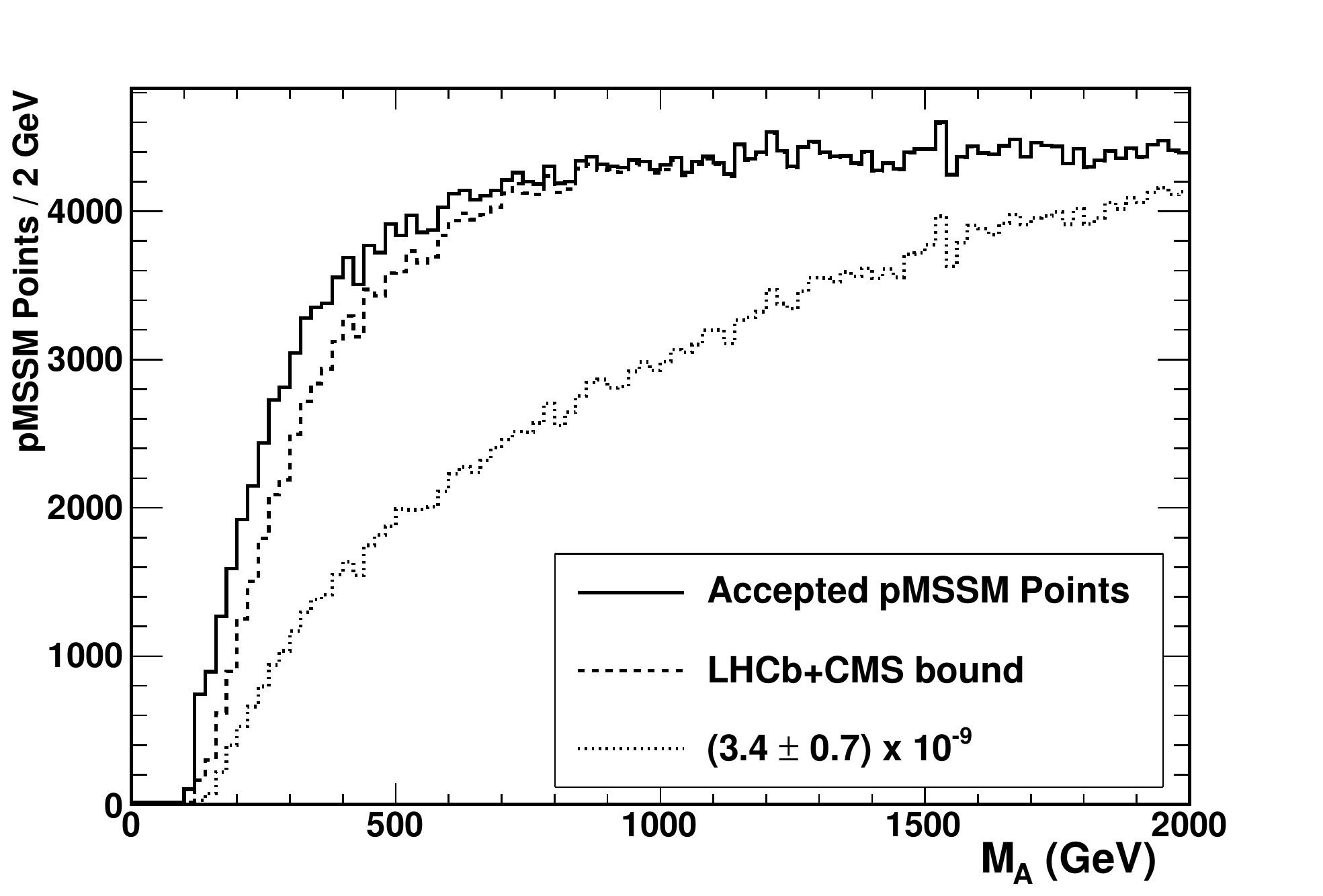} \\
\includegraphics[width=0.35\textwidth]{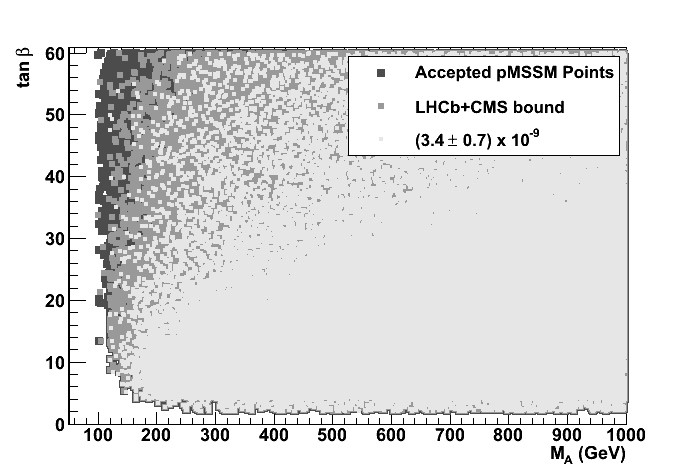} \\
\end{tabular}
\end{center}
\caption{Distribution of pMSSM points after the $B_s \rightarrow \mu^+ \mu^-$ constraint
projected on the $M_A$ (upper panel) and ($M_A , \tan \beta$) plane (lower panel) for all accepted 
pMSSM points (medium grey), points not excluded by the combination of the present LHCb and CMS 
analyses (dark grey) and the projection for the points compatible with the measurement of 
the SM expected branching fractions with a 20\% total uncertainty (light grey).}
\label{fig:bsmumu}
\end{figure}

We compare our accepted pMSSM points to this limit as well as to the projected constraint 
in the case of observation of the decay with a SM-like rate of BR($B_s\to\mu^+\mu^-$) = 
(3.4 $\pm$ 0.7)$\times$10$^{-9}$, to which we have attached a 20\% total relative uncertainty, 
by the end of the 2012 run. The results are presented in Figure~\ref{fig:bsmumu} in the $M_A$ 
and $M_A$--$\tan \beta$ planes. The current limit affects the low $M_A$ values up to 700~GeV, 
excluding large $\tan\beta$ values, below $M_A \sim$200~GeV. The projected constraint has a stronger 
impact, with more than half of the spectrum being excluded for $M_A\lesssim 700$ GeV. However, 
the low $\tan\beta$ region at $\tan\beta \sim 5$ remains largely unaffected also by this constraint. 

\subsection{Dark matter direct detection experiments}
\label{sec:3-4}

Dark matter direct detection experiments have made great progress exploring $\tilde\chi p$ scattering 
cross sections in the range predicted by the MSSM~\cite{Ahmed:2009zw,Aprile:2011hi}. In particular, the 
recent XENON~100 result~\cite{Aprile:2011hi}, places a 90\% C.L. upper bound on the spin-independent 
$\tilde\chi p$ cross section around 10$^{-8}$~pb for $M_{\mathrm{WIMP}} \simeq$ 100~GeV and excludes 
$\simeq$20\% of the accepted pMSSM points in our scan. 
By the end of 2012, this bound should be improved by a factor of 7, if no signal is observed, 
which will exclude $\simeq$50\% of the accepted pMSSM scan points within our scan range.
\begin{figure}[t!]
\begin{center}
\begin{tabular}{c}
\includegraphics[width=0.35\textwidth]{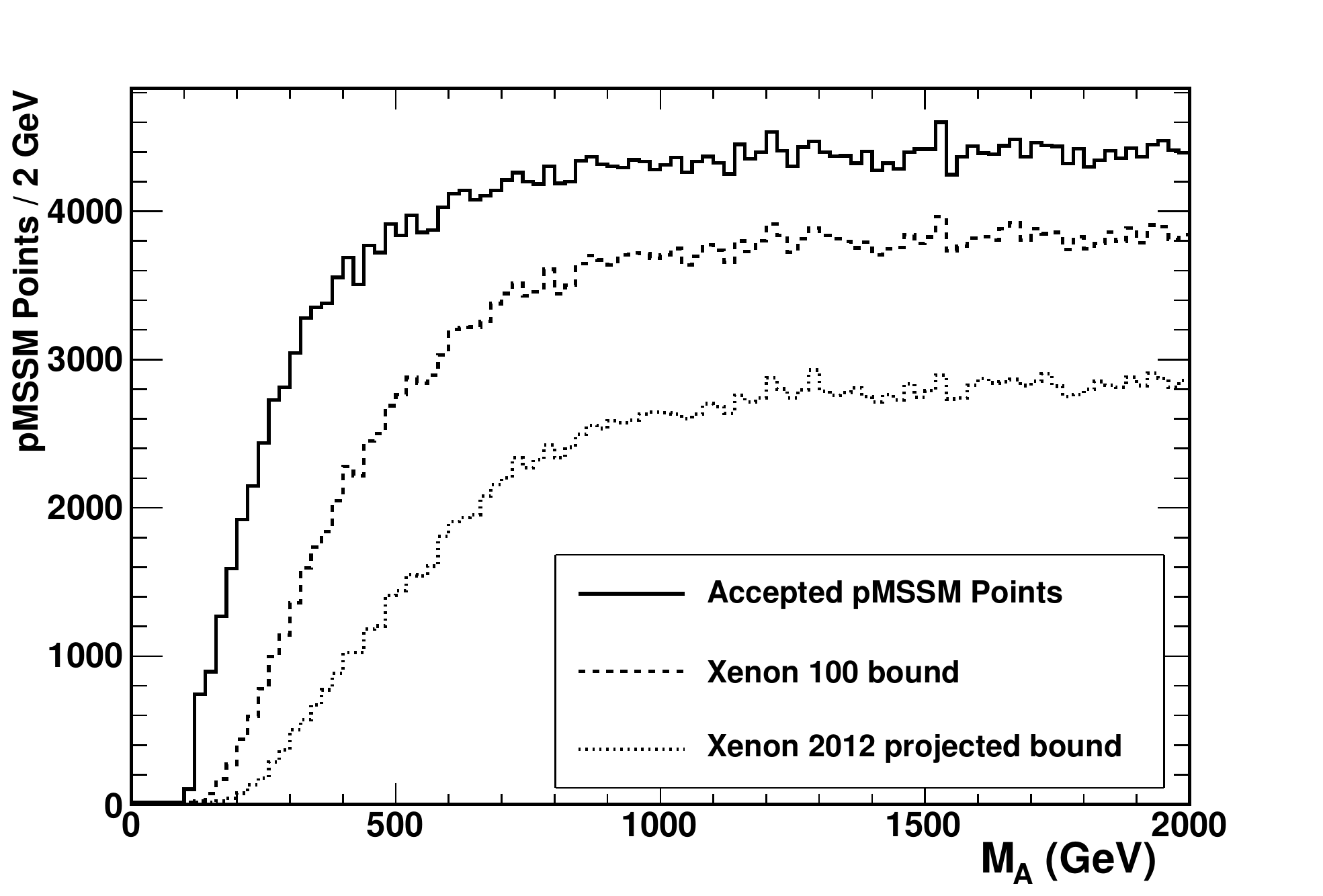} \\
\includegraphics[width=0.35\textwidth]{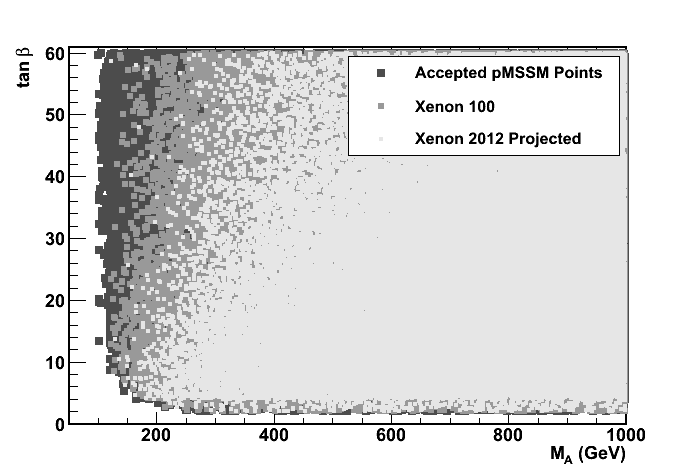} \\
\end{tabular}
\end{center}
\caption{Distribution of pMSSM points after the dark matter direct detection constraint projected on the 
$M_A$ (upper panel) and ($M_A , \tan \beta$) plane (lower panel) for all accepted pMSSM points (medium grey), 
points not excluded by the current XENON-100 data (dark grey) and the projection for the XENON sensitivity 
at the end of 2012 (light grey).}
\label{fig:DD}
\end{figure}
The $\tilde\chi p$ spin-independent scattering process has contributions from scalar quark exchange and 
t-channel Higgs exchange~\cite{Jungman:1995df}. 
The latter dominates over vast region of the parameter space with the Higgs coupling to the proton depending 
on its coupling to gluons, through a heavy quark loop and to non-valence quarks. The scattering cross section 
retains a strong sensitivity on the CP-odd boson mass as highlighted in Figure~\ref{fig:DD} which shows the pMSSM 
points retained after the XENON~100 and the projected 2012 sensitivity. The 2012 data should exclude virtually 
all solutions with $M_A \lesssim$ 200~GeV, independent on the value of $\tan \beta$, if no signal is detected.

\section{The $h^{0}$ boson at LHC and MSSM constraints}
\label{sec:4}

The ATLAS and CMS experiments have recently combined the results of their searches for a SM-like Higgs 
boson on 1.0-2.3~fb$^{-1}$ of statistics per experiment~\cite{CMS_plus_ATLAS} obtaining a 95\% C.L.\ upper 
limit on the ratio $R_{ff} = \frac{\sigma \times \mathrm{BR(h^0 \to ff)}}{\sigma_{\mathrm{SM}} \times 
\mathrm{BR(H^0_{SM} \to ff)}}$ of 2.23 at $M_H$ = 120~GeV. Updated preliminary results of the individual 
searches based on almost 5~fb$^{-1}$  
of statistics, which have just been released, are sensitive to $R \simeq$ 1 and give possible hints of 
an excess in the number of events recorded by both experiments in several channels around a mass value of 
$\sim$125~GeV~\cite{seminar,ATLAS-CONF-2011-163,11-032}. 
Here, we consider two distinct scenarios suggested by the 2011 LHC data and the constraints derived on the 
MSSM. First we consider the effect of the observation of a Higgs boson with a mass in the range 
123 $< M_h <$ 127~GeV, which corresponds to the case the reported excess would be confirmed by the 2012 data
and where the upper end of the range represents the current 95\%~C.L.\ upper limit from LHC. 
 
Alternatively, we study the scenarios with a significant suppression of the Higgs boson yield compared to 
the SM, corresponding to the exclusion of a SM-like Higgs boson.  In both cases, the bounds on 
$M_A$, $\tan \beta$ and strongly interacting sparticles define the available parameter space for studying 
the $h^0$ mass and the suppression of its couplings and thus the reduction of the yields in the LHC Higgs 
searches. We consider the two processes $gg \rightarrow h^0 \to \gamma \gamma$ and 
$gg \rightarrow h^0 \to W^+ W^-$, $Z^0Z^0$, which have the largest sensitivities in the present searches at 
the LHC.

\subsection{Constraints from $M_h$ determination}
\label{sec:4-1}

The determination of the mass of the lightest Higgs boson with an accuracy of order of 1~GeV 
places some significant constraints on the SUSY parameters, in particular in the typical mixing scenario,
where its central value corresponds to a mass close to the edge of the range predicted in the MSSM. 
In order to evaluate these constraints, we select the accepted pMSSM points from our scans, which have 
123 $< M_{h^0} <$ 127~GeV. These are $\simeq$20\% of the points not already excluded by the LHC SUSY searches 
with 1~fb$^{-1}$ in our scan, where parameters are varied in the range given in the central column of 
Table~\ref{tab:paramSUSY}.
\begin{figure}[h!]
\begin{center}
\includegraphics[width=0.4\textwidth]{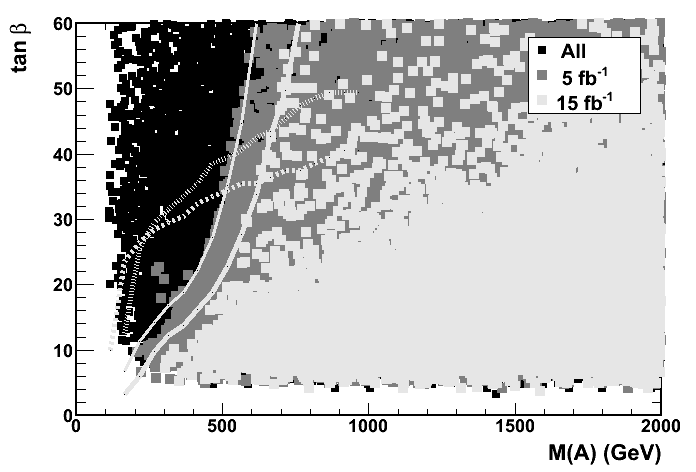} \\
\includegraphics[width=0.4\textwidth]{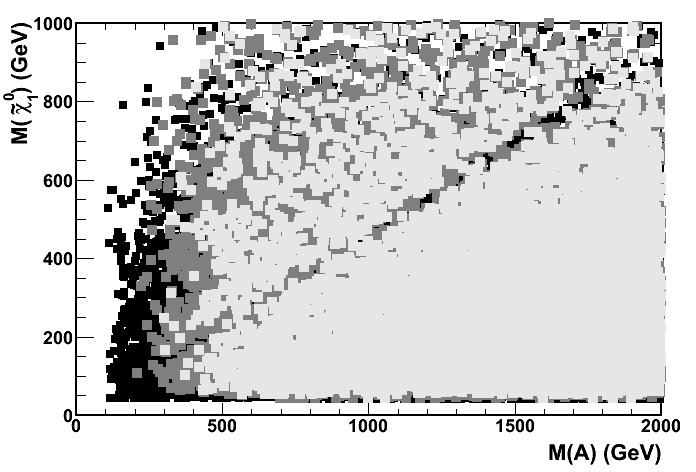} \\
\includegraphics[width=0.4\textwidth]{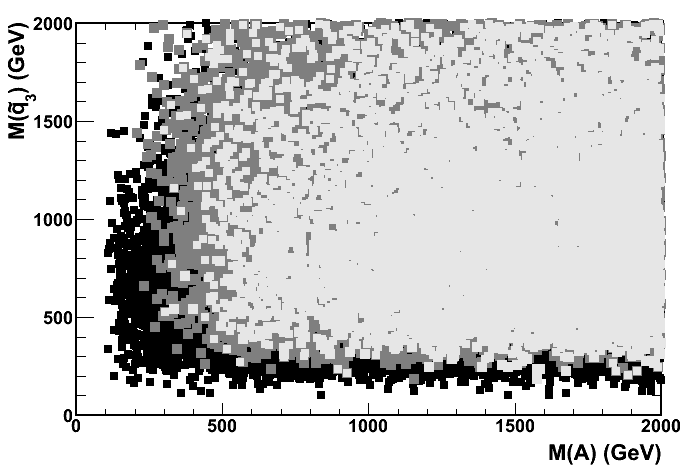} \\
\end{center}
\caption{pMSSM points in the ($M_A$, $\tan \beta$) (top panel), ($M_A$, $M_{\tilde \chi^0_1}$) 
(centre panel) and ($M_A$, $M_{\tilde q_3}$), where $M_{\tilde q_3}$ is the minimum of the masses 
of the $\tilde t_1$ and $\tilde b_1$ squarks, (bottom panel) parameter space, giving 123 $< M_H <$ 
127~GeV. The different shades of grey show all the valid pMSSM points without cuts (black) and those fulfilling 
the Higgs mass cut allowed by the 2011 data (dark grey) and by the projected 2012 data (light grey), 
assuming no signal beyond the lightest Higgs boson is observed. The lines in the top plot show the regions 
which include 90\% of the scan points 
for the $A \rightarrow \tau^+ \tau^-$ and $B_s \rightarrow \mu^+ \mu^-$ decays at the LHC and the dark 
matter direct detection at the XENON experiment, for the caption see Figure~\ref{fig:matanb}. The narrow 
corridor along the diagonal in the ($M_A$, $M_{\tilde \chi^0_1}$) plane corresponds to the $A^0$ funnel region 
where the $\chi \chi \to A$ annihilation reduces $\Omega_{\chi} h^2$ below the accepted range.}
\label{fig:matanb125}
\end{figure}

\begin{figure}[h!]
\begin{center}
\includegraphics[width=0.4\textwidth]{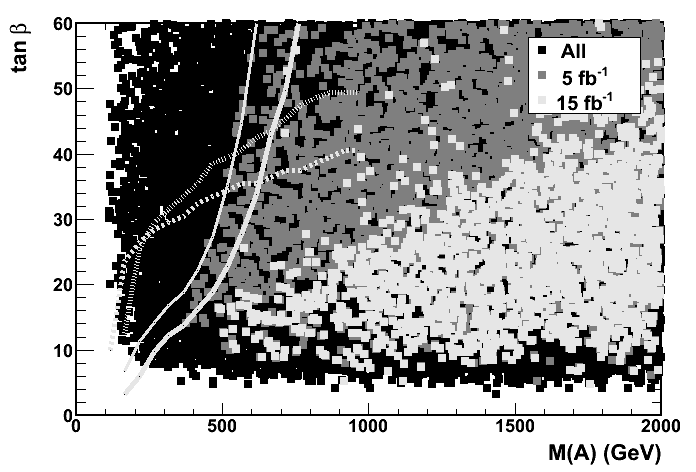} \\
\includegraphics[width=0.4\textwidth]{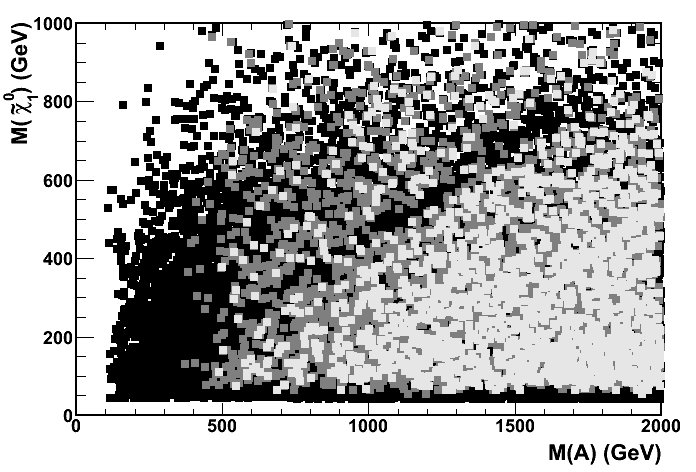} \\
\includegraphics[width=0.4\textwidth]{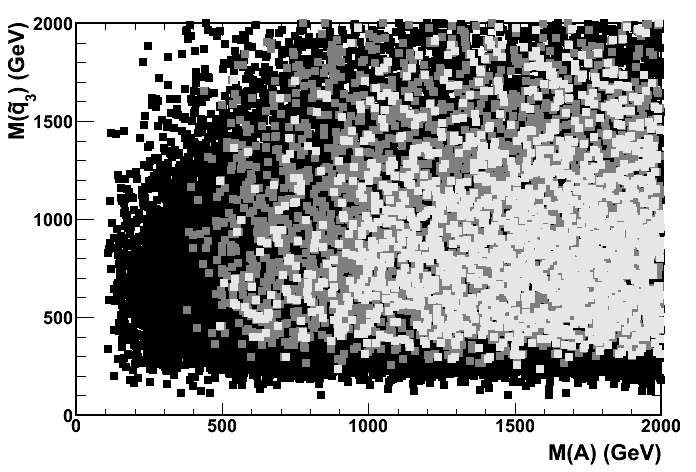} \\
\end{center}
\caption{pMSSM points in the parameter planes ($M_A$, $\tan \beta$) (top panel), ($M_A$, $M_{\tilde \chi^0_1}$) 
(centre panel) and ($M_A$, $M_{\tilde q_3}$) (bottom panel), where $M_{\tilde q_3}$ is the minimum 
of the masses of the $\tilde t_1$ and $\tilde b_1$ squarks, giving 123 $< M_{h} <$ 
127~GeV, after imposing the additional requirements on the Higgs rates. The color coding is the same 
as for Figure~\ref{fig:matanb125}.}
\label{fig:matanb125RH}
\end{figure}
\begin{figure}[h!]
\begin{center}
\begin{tabular}{c c}
\includegraphics[width=0.23\textwidth]{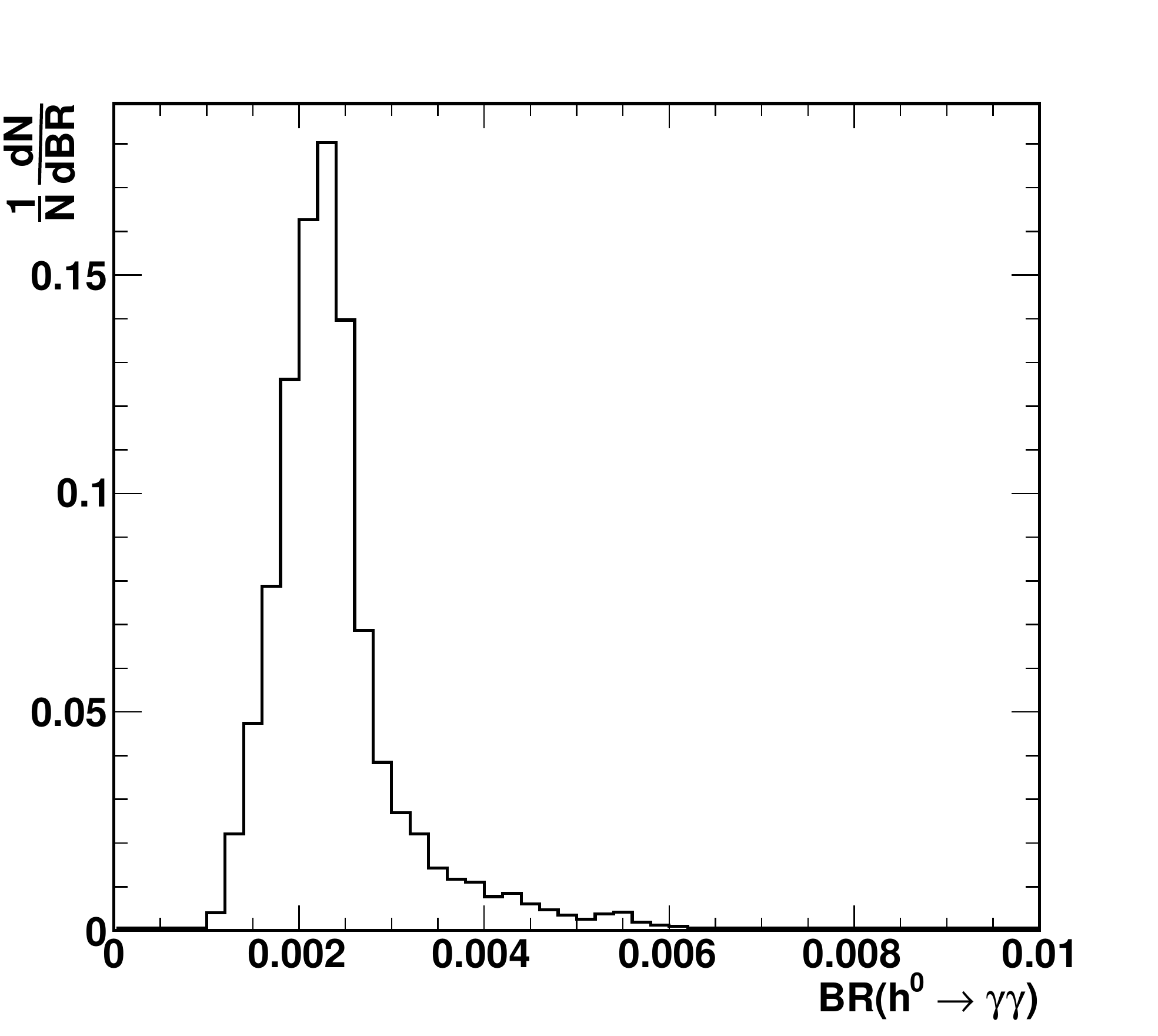} &
\includegraphics[width=0.23\textwidth]{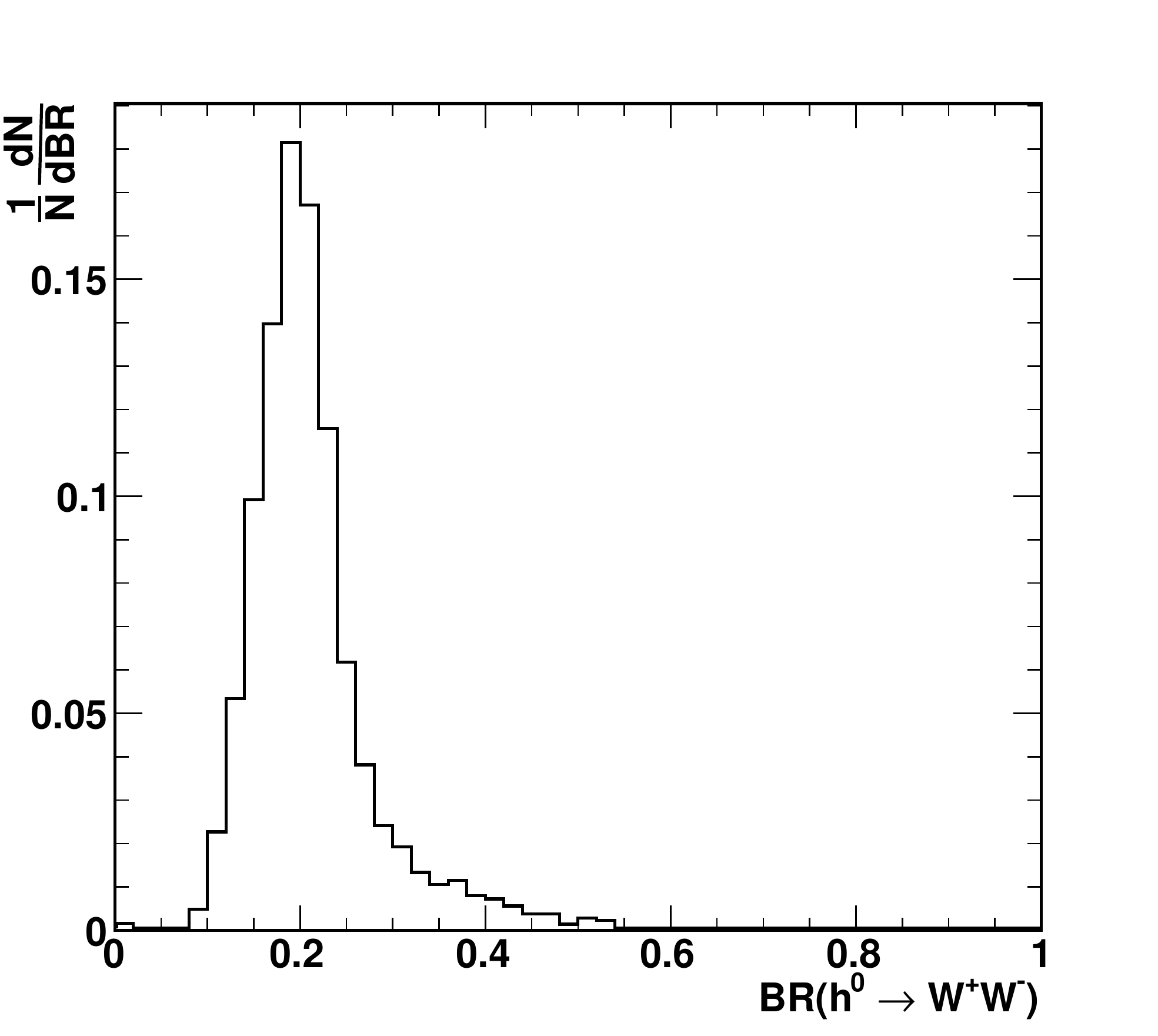} \\
\end{tabular}
\end{center}
\caption{Branching fractions for $h^0$ decays into $\gamma \gamma$ (left panel) and $W^+W^-$ (right panel) 
pairs calculated with {\tt HDECAY} for selected pMSSM points fulfilling the 2011 data constraints.}
\label{fig:br}
\end{figure}

Figure~\ref{fig:matanb125} shows the points fulfilling these conditions, which are also allowed by the other 
2011 data constraints and by the 2012 projection. The parameter space is 
defined by three combinations of variables: $M_A$ -- $\tan \beta$, $M_A$ -- $M_{\tilde \chi^0_1}$ 
and $M_A$ -- $M_{\tilde q_3}$, where $M_{\tilde q_3}$ is the minimum of the masses of the $\tilde t_1$ 
and $\tilde b_1$ squarks. We observe that imposing the value of $M_{h^0}$ selects a broad wedge in the 
($M_A$, $\tan \beta$)  plane, at rather heavy $A^0$ masses and moderate to large values of $\tan \beta$ and 
extending beyond the projected sensitivity of the searches in the $A^0 \to \tau^+ \tau^-$ but also that of 
direct DM detection and would be compatible with a SM-like value for the rate of the $B_s^0 \to \mu^+ \mu^-$ 
decay. Next, we impose the condition that the yields in the $\gamma \gamma$, $W^+W^-$ and $Z^0Z^0$ final 
states reproduce the observed rates of candidate events reported by the ATLAS and CMS collaborations. We require 
that 1$\le R_{\gamma \gamma} <$3 and 0.3$< R_{W^+W^-/Z^0Z^0} <$2.5.
The points fulfilling these constraints are shown in Figure~\ref{fig:matanb125RH}. Here, we observe that the wedge 
in the ($M_A$, $\tan \beta$) plane is further restricted and solutions with $M_{\chi^0_1} > M_A$ are also strongly 
suppressed. The branching fractions for $h^0 \to \gamma \gamma$ and $h^0 \to W^+W^-$, obtained with the HDECAY 
program, for the selected pMSSM points fulfilling the above mentioned constraints and with 123 $ < M_h <$ 127~GeV 
are shown in Figure~\ref{fig:br}. It is interesting to observe that a rather broad range of values are possible, depending 
on the SUSY parameters, but the ratios $R_{\gamma \gamma}$ and $R_{WW}$ are always highly correlated. In our pMSSM 
scans, we do not find solution where the $\gamma \gamma$ yield is significantly enhanced compared to the SM while those 
in $WW$ and $ZZ$ are either unchanged or suppressed. 
\begin{figure}[h!]
\begin{center}
\begin{tabular}{c c}
\includegraphics[width=0.23\textwidth]{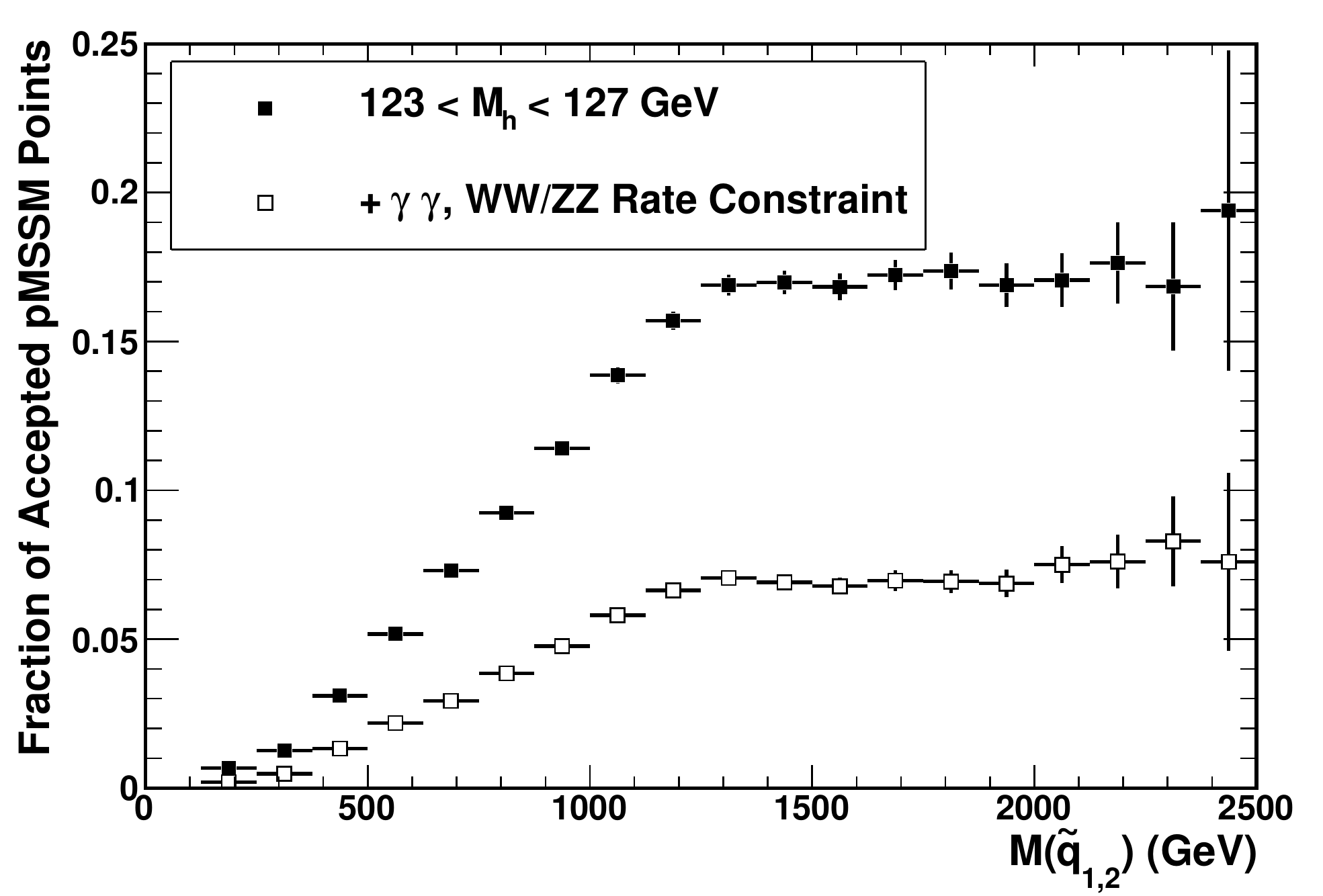} &
\includegraphics[width=0.23\textwidth]{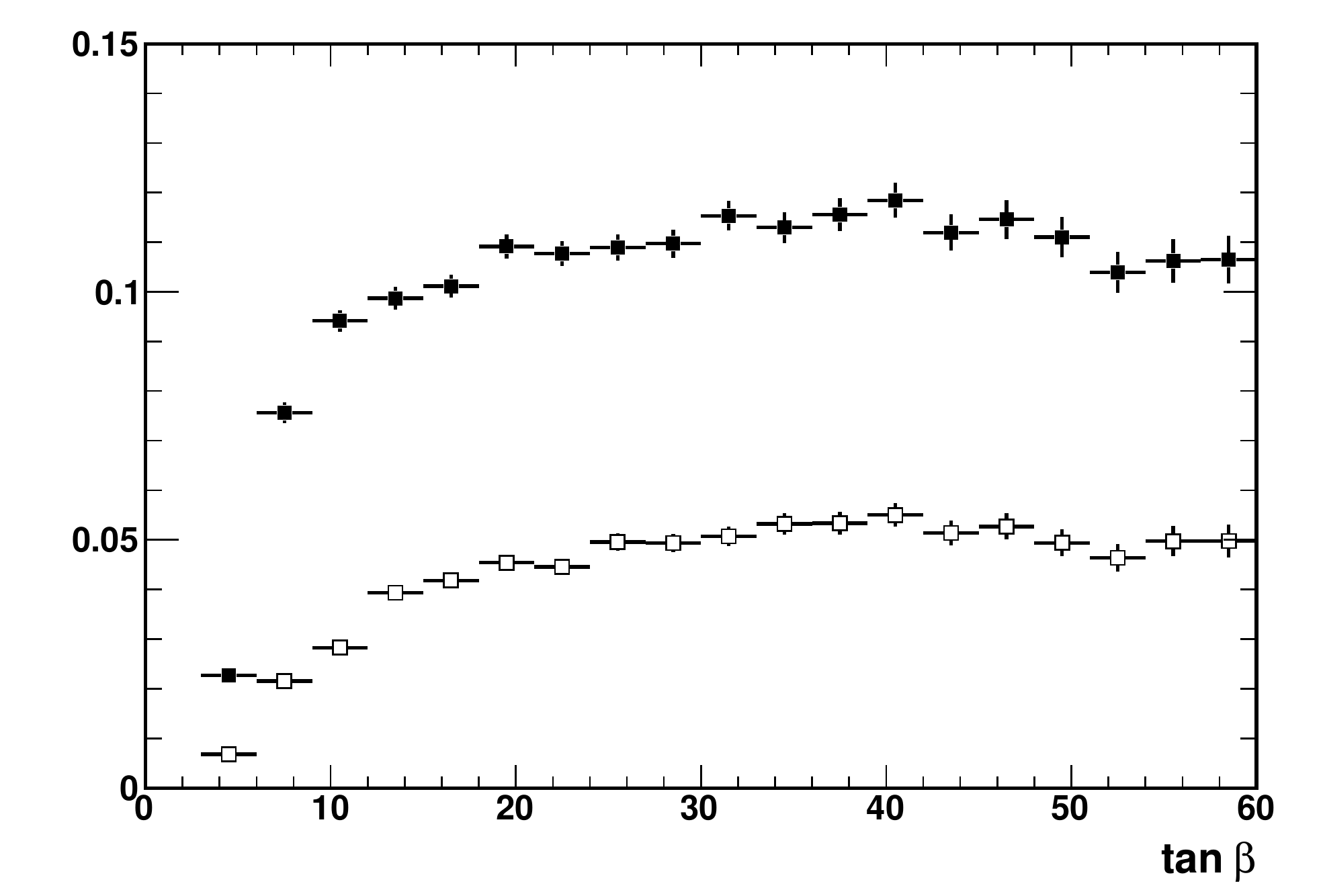} \\
\includegraphics[width=0.23\textwidth]{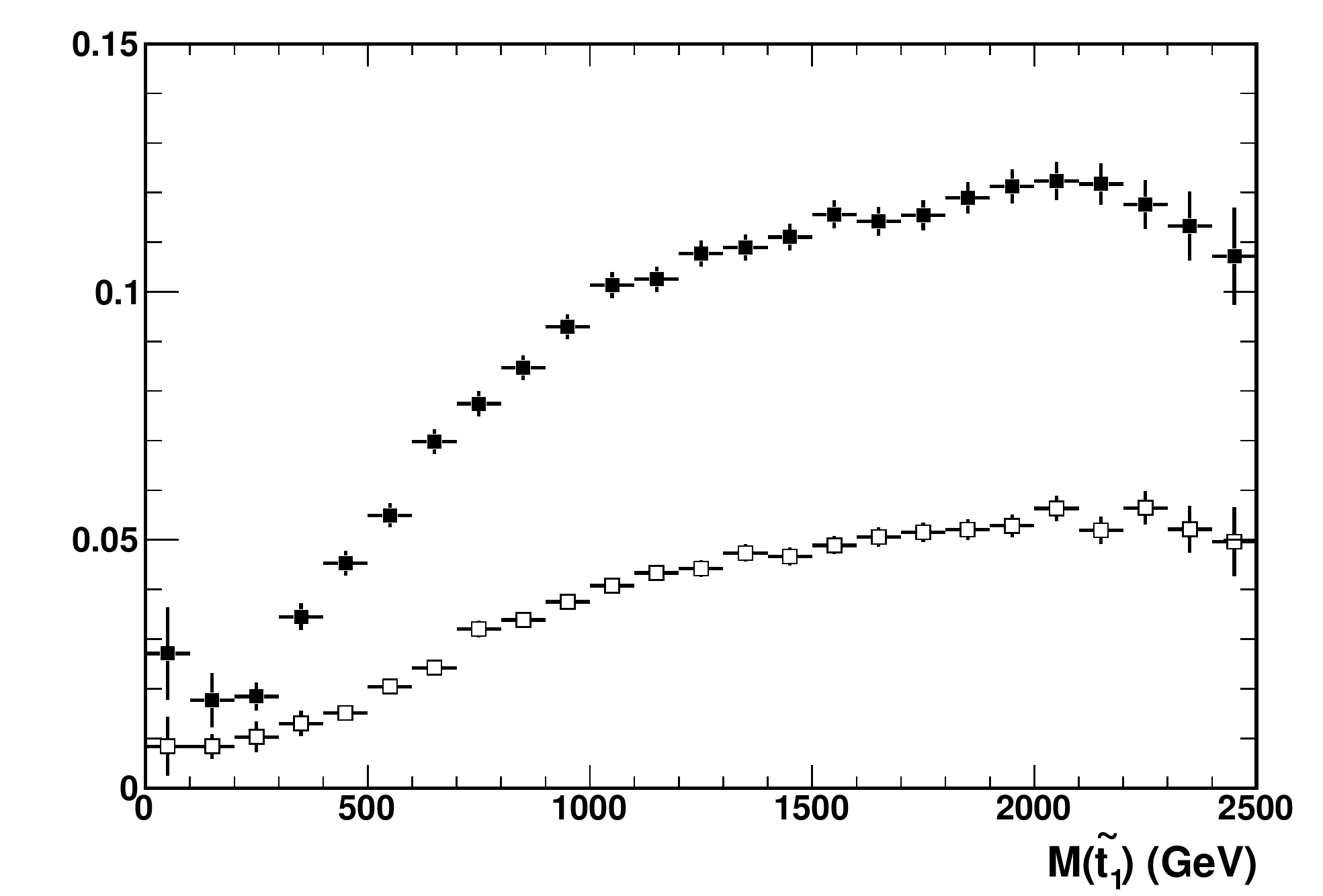} &
\includegraphics[width=0.23\textwidth]{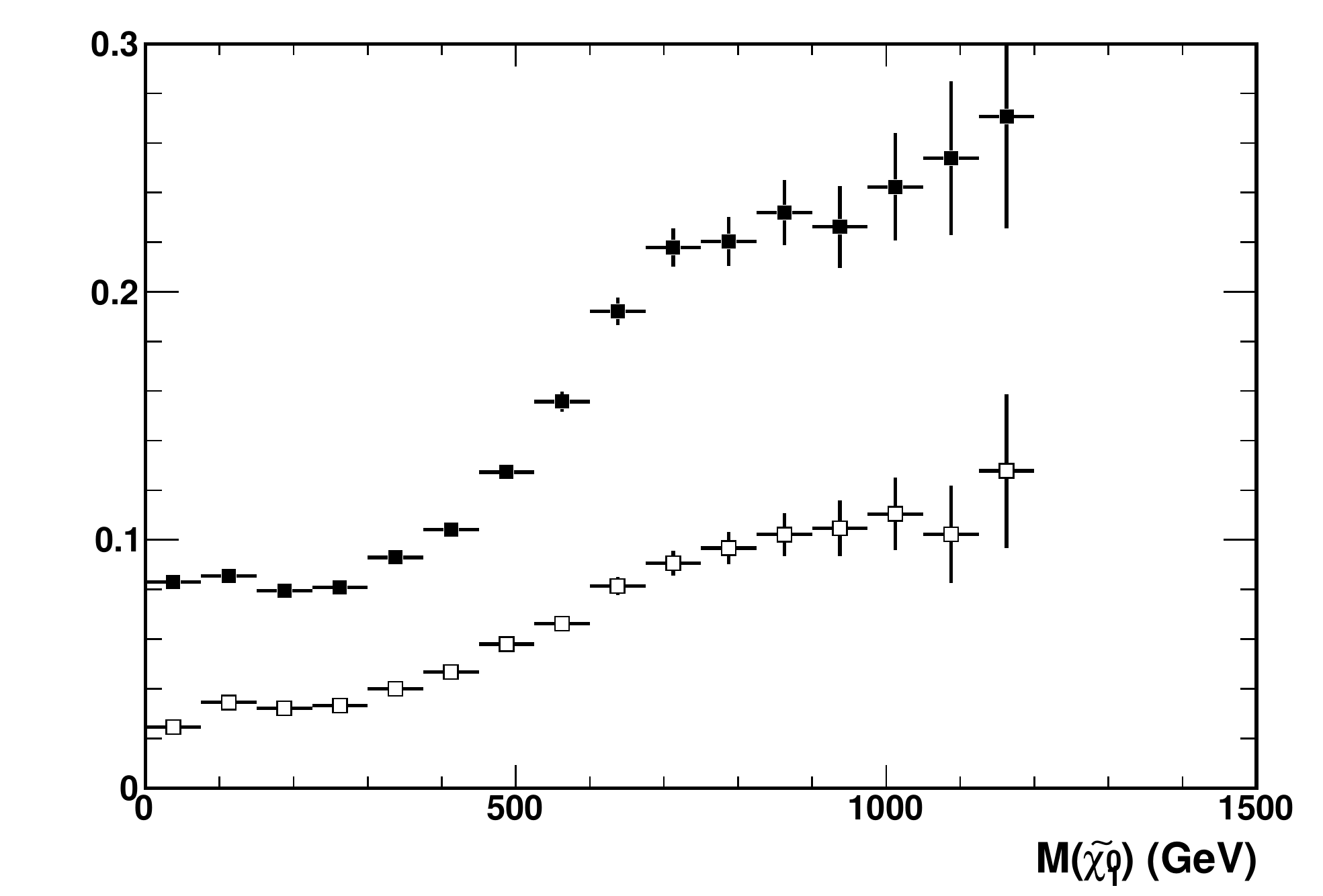} \\
\end{tabular}
\end{center}
\caption{Fraction of accepted pMSSM points, with 123 $< M_{h}<$ 127~GeV (filled squares), 
not excluded by the SUSY searches with 15~fb$^{-1}$ of 7~TeV data as a function of the mass of the lightest 
squark of the first two generations (upper left panel), of the mass of the scalar top $\tilde t_1$ (lower left 
panel), of $\tan \beta$ (upper right panel) and of the lightest neutralino $\tilde \chi^0_1$ (lower right panel). 
The open square points show the fraction of pMSSM points after imposing the additional requirements on the Higgs rates.}
\label{fig:msq125}
\end{figure}
The effect of the Higgs constraints on some pMSSM parameters is shown in Figure~\ref{fig:msq125}, in terms of the 
fraction of valid pMSSM points, fulfilling the general requirements discussed in Section~\ref{sec:2}, those from 
searches by the end of 2012 and giving  123 $< M_h <$ 127~GeV. 
In particular, a comparison of Figure~\ref{fig:msq125} with Figure~\ref{fig:st1b1}, which differ for the 
requirements on $M_h$,  
shows that values of $\tan \beta \le 6$ become disfavoured, while the masses of scalar quarks are not 
significantly affected. Imposing the condition that the yields in the $\gamma \gamma$, $W^+W^-$ and $Z^0Z^0$ 
final states are consistent with the observed rates of candidate Higgs events reduces the fraction of accepted 
points preferentially at large masses of $\tilde t_1$, $\tilde q_{1,2}$ and $\tilde \chi^0_1$.
 
In order to estimate the effect of the program used for computing the $h^0$ mass and decay branching fractions,  
we repeat the analysis using {\tt FeynHiggs} and compare the results. We observe that 20.1\% and 25.2\% of the 
accepted pMSSM points in our scan have Higgs mass in the range 123 $< M_h <$ 127~GeV using {\tt SoftSUSY} 
and {\tt FeynHiggs}, respectively. Of these 12\% have $R_{\gamma \gamma} \ge$ 1 using 
{\tt HDECAY} and 7.2\% using {\tt FeynHiggs}.
Finally, we compute the fine tuning parameter $\Delta$, using the definition of ref.~\cite{Perelstein:2007nx}, 
for the points in the accepted Higgs mass range and for those having also the $\gamma$, $WW$ and $ZZ$ rates 
within the constraints used above, and find that 20.6\% and 18.4\% of them have $\Delta <$ 100.

\subsection{Constraints from no Higgs observation}
\label{sec:4-2}

In this section, we turn to study the implications of a non-observation of the Higgs boson at the LHC.
We select pMSSM points compatible with the current constraints and such that no signal beyond the 
SM is observed in the 2011 and 2012 statistics at LHC and XENON. Further, we request a rate suppression 
resulting in $R_{\gamma \gamma}$ and $R_{WW}$, $R_{ZZ}$, the ratio of the product $\sigma \times \mathrm{BR}$
to its SM expectations, $\le$ 0.3  in the $\gamma \gamma$ and $WW$, $ZZ$ channels. Such a suppression can 
be considered significant for the perspectives to discover, or exclude, a Higgs boson by the end of 2012. 

\begin{figure}[h!]
\begin{center}
\includegraphics[width=0.4\textwidth]{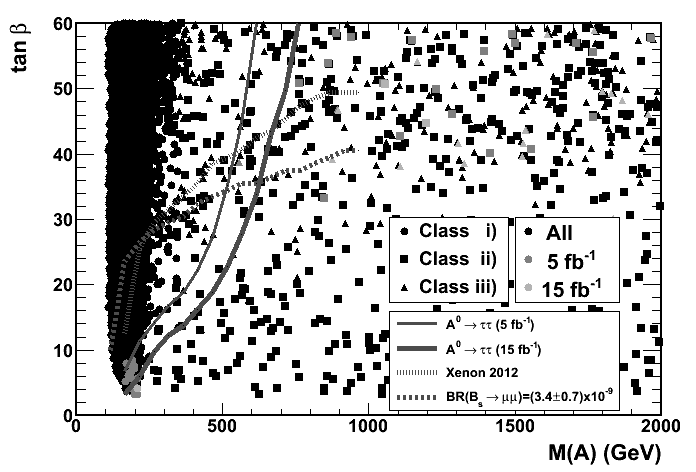} \\
\includegraphics[width=0.4\textwidth]{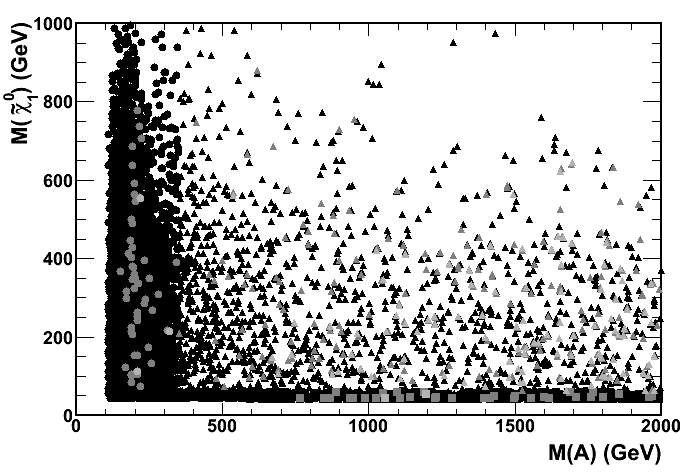} \\
\includegraphics[width=0.4\textwidth]{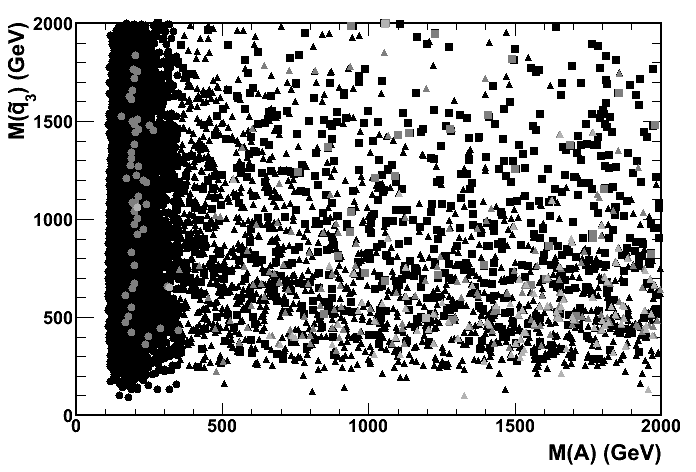} \\
\end{center}
\caption{pMSSM points in the ($M_A$, $\tan \beta$) (top panel), ($M_A$, $M_{\tilde \chi^0_1}$) 
(centre panel) and ($M_A$, $M_{\tilde q_3}$), where $M_{\tilde q_3}$ is the minimum of the masses of 
the $\tilde t_1$ and $\tilde b_1$ squarks (bottom panel) parameter space, giving a suppression of the 
$gg \to h^0 \to \gamma \gamma$ rate compared to the SM prediction, corresponding to $R_{\gamma \gamma} \le$ 0.3. 
The different shades of grey show the 
points allowed in the pMSSM without cuts and those allowed by the 2011 data and by the projected 2012 
data, assuming no signal is observed. The point shape gives the classification of the points depending 
on the relevant suppression mechanism, as discussed in the text. The lines in the top plot show the regions 
which include 90\% of the scan points for the $A \rightarrow \tau^+ \tau^-$ and $B_s \rightarrow \mu^+ \mu^-$ 
decays at the LHC and the dark matter direct detection at the XENON experiment.}
\label{fig:matanb}
\end{figure}
\begin{figure}[h!]
\begin{center}
\includegraphics[width=0.4\textwidth]{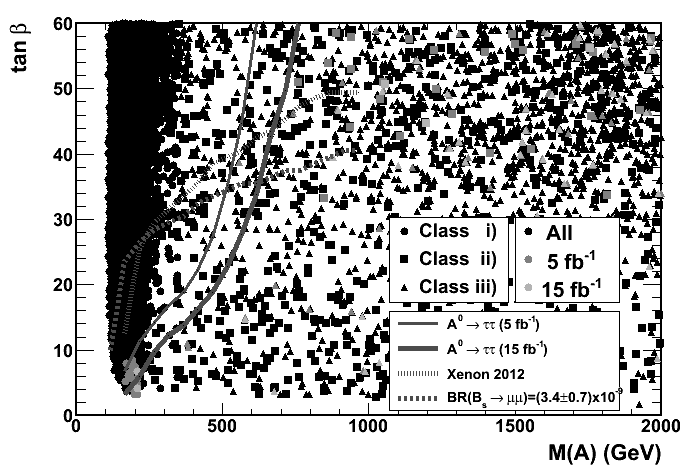} \\
\includegraphics[width=0.4\textwidth]{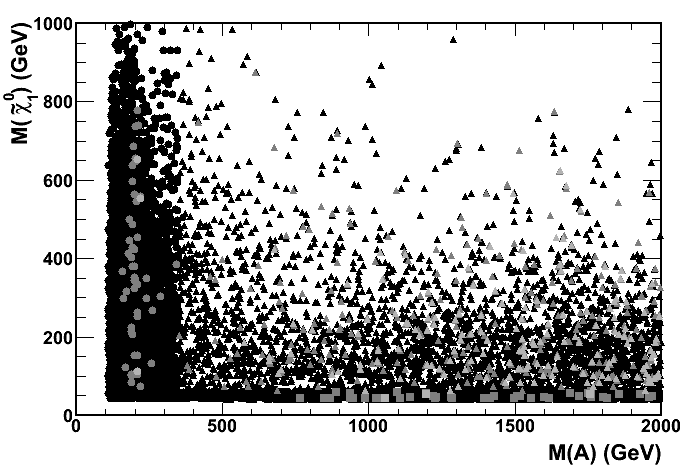} \\
\includegraphics[width=0.4\textwidth]{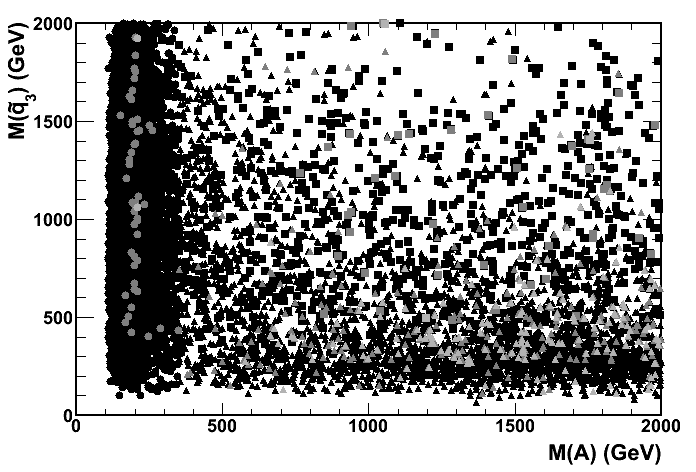} \\
\end{center}
\caption{pMSSM points in the ($M_A$, $\tan \beta$) (top panel), ($M_A$, $M_{\tilde \chi^0_1}$) 
(centre panel) and ($M_A$, $M_{\tilde q_3}$), where $M_{\tilde q_3}$ is the minimum of the masses of 
the $\tilde t_1$ and $\tilde b_1$ squarks (bottom panel) parameter space, giving $R_{WW}$, $R_{ZZ} <$ 0.3. 
The symbol colour and shape coding is the same as in Figure~\ref{fig:matanb}.} 
\label{fig:matanbW}
\end{figure}

Analysing the pMSSM points fulfilling this selection, we observe that three distinct classes of 
scenarios emerge: i) the region of the non-decoupling scenario 
with $M_A <$ 250~GeV and $\tan \beta \sim$ 5 , ii)  the invisible Higgs scenario with 
$M_{\tilde\chi^0_1} < M_{h^0}$ and small $|\mu|$ and iii) the region with light 
$\tilde{t}_1$, $\tilde{b}_1$ squarks. These three scenarios are realised, respectively, in $\sim$10$^{-2}$, 
5$\times$10$^{-4}$,and 4$\times$10$^{-4}$ of the valid pMSSM points in our scans.
The evolution of the parameter space for these scenarios after applying the constraints for 5~fb$^{-1}$ of 
LHC data plus the current XENON~100 limit and the projected constraints for 2012 for 15~fb$^{-1}$ of LHC 
data and the forthcoming XENON run are summarised in Figures~\ref{fig:matanb} and \ref{fig:matanbW}, 
for $\gamma \gamma$ and $W^+W^-$, $Z^0 Z^0$, using the combinations of parameters as in the previous section.

\begin{figure}[h!]
\begin{center}
\includegraphics[width=0.4\textwidth]{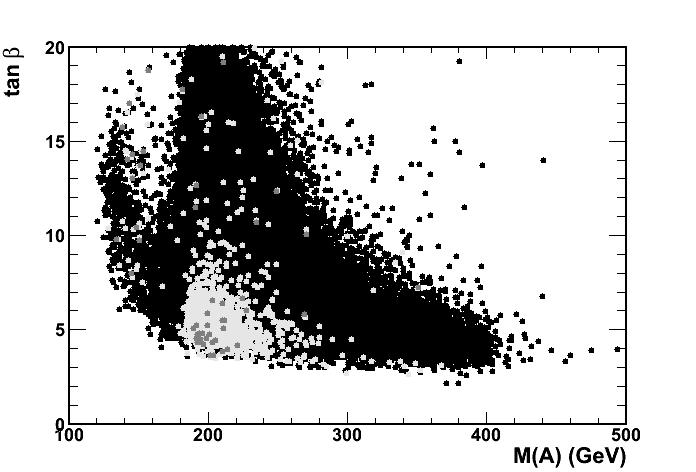}
\end{center}
\caption{pMSSM points in the ($M_A$, $\tan \beta$) plane with rate for $bb \to H^0$/$A^0$, $gg \to H^0$/$A^0$; 
$H^0$/$A^0 \to Z^0 Z^0$ non-zero (black points) and exceeding the expected sensitivity to $Z^0Z^0$ final 
states from the $H^0_{SM}$ search for 5 (medium grey) and 15~fb$^{-1}$ (light grey) of integrated luminosity.}
\label{fig:HZZ}
\end{figure}

Class i) represents the tiny part of the low $M_A$ and $\tan \beta$ region left by $A^0 \to \tau \tau$ 
and DM direct detection. Only $\sim$8$\times$10$^{-2}$ and 7$\times$10$^{-4}$ of the points in these region 
remain viable after applying the anticipated constraints from the LHC and XENON for the 2011 and 2012 data, 
respectively. In this region the decay $H^0 \to Z^0 Z^0$, leading to the same final state 
as the SM Higgs process $H^0_{SM} \rightarrow Z^0 Z^0$~\cite{Aad:2011uq,11-015}, is open. The results of the 
intermediate mass SM Higgs can be re-interpreted to place constraints on the low $M_A$ scenario. We compare 
the product of cross section and decay branching fraction for $gg \rightarrow H^0/A^0 \to Z^0 Z^0$ to the LHC 
sensitivity for the SM Higgs decay into $Z^0Z^0$, rescaled by the assumed integrated luminosity. 
Figure~\ref{fig:HZZ} shows the regions of the ($M_A$, $\tan \beta$) plane where the $H^0 \to Z^0 Z^0$ rate 
exceeds the LHC sensitivity for 5 and 15~fb$^{-1}$. These cover part of the scenario i) parameter space 
left after the other constraints.

The light neutralino scenario (class ii)) may prove difficult to test directly in the squark and gluino sector 
due to the small transverse energy released. The dominant invisible Higgs decay, responsible for the yield 
suppression in the canonical channels, represents a distinct signature which should be feasible to test 
experimentally~\cite{Kinnunen:2005aq}. From our simulation we estimate that $\sim$7$\times$10$^{-2}$ of the 
points in this scenario should remain not excluded, after applying the constraints from the 2012 LHC and direct DM 
searches. Finally, for class iii) dedicated searches for stop and sbottom production and decay, involving $t$ and 
$b$-tagging in events with jets and $MET$ are already under way~\cite{ATLAS:2011bmet,CMS:2011bmet}, and will be 
further pursued to probe the mass region relevant for a possible Higgs rate suppression.

\section{Conclusion}
\label{sec:5}

The Higgs boson searches at the LHC, in conjunction with those for $B^0_s \rightarrow \mu^+ \mu^-$ 
again at the LHC and dark matter direct detection in underground experiments, place highly constraining 
bounds on the MSSM parameters. Inspired by the preliminary results reported by the ATLAS, CMS and LHCb 
collaborations, we have analysed two scenarios. The first has a light Higgs boson, with mass 123 $< M_h <$ 
127~GeV, no signal from the $MET$ searches and the $B^0_s \rightarrow \mu^+ \mu^-$ decay with SM-like 
branching fraction. The second has no light Higgs boson within a factor of three from the SM rate and again 
no signal from strongly interacting sparticles and SM-like $B^0_s \rightarrow \mu^+ \mu^-$ decay rate. 

We perform flat scans of the 19-parameter pMSSM space imposing constraints from searches at LEP and the 
Tevatron, flavour physics and dark matter relic density. We observe that imposing the mass of the lightest 
Higgs boson in the range 123 $< M_h <$ 127~GeV restricts the parameter space within a wedge in the 
($M_A$, $\tan \beta$) plane, corresponding to rather large values of the $A^0$ mass and moderate to large values 
of $\tan \beta$, while it does not significantly affect the values of the masses of weakly interacting 
supersymmetric particle partners. Further imposing that the yields in the $\gamma \gamma$, $W^+W^-$ and $Z^0Z^0$ 
final states reproduce the rates of candidate events reported by the ATLAS and CMS collaborations the wedge 
in the ($M_A$, $\tan \beta$) plane becomes more pronounced and the fraction of accepted points gets reduced 
preferentially at large masses of $\tilde t_1$, $\tilde q_{1,2}$ and $\tilde \chi^0_1$.

On the contrary, a non-observation of the Higgs boson, corresponding to a suppression of its yields in the 
$\gamma \gamma$, $WW$ and $ZZ$ final states to a factor of $\simeq$3 compared to the SM expectations, would 
confine the viable sets of MSSM parameters to just three narrow regions having light $A^0$ and low 
$\tan \beta$ values, $M_{\tilde \chi^0_1} < M_h^0$ and invisible $h^0$ decays or light $\tilde t_1$, 
$\tilde b_1$ scalar quarks. Each of these scenarios consists of a tiny fraction of the valid pMSSM points and 
can be independently probed, through a direct search for the $A^0$ and $H^0$ Higgs states, light neutralinos 
and invisible Higgs decays and light to intermediate mass scalar top quarks. 
While it is still too early to draw definite conclusions from 
the preliminary LHC results, the continuation of all these searches at the LHC and dark matter experiments 
should provide us with a definite test of the MSSM independent on the mass scales of the scalar quarks of the 
first two generations and of the gluino.

\section*{Acknowledgements}
We would like to thank M.~Mangano for supporting this activity and the LPCC for making dedicated  
computing resources available to us. We are grateful to Abdelhak Djouadi for extensive discussion 
and his suggestions. We also acknowledge useful discussions with G.~Belanger, M.~Spira 
and S.~Heinemeyer. Several colleagues in the LHC collaborations provided us with valuable feedback, 
in particular A.~De Roeck and B.~Hooberman. We are grateful to E.~Gianolio for computing support 
and to E.~Aprile, P.~Beltrame and A.~Melgarejo for providing us with information on the current 
and expected sensitivity of the XENON experiment. 


\bibliographystyle{epjc}
\bibliography{higgs-lhc}

\end{document}